\documentclass[12pt]{article}
\pdfoutput=1
\usepackage[margin=3cm]{geometry}
\usepackage[utf8]{inputenc}
\usepackage{amsmath}
\usepackage{amssymb}
\usepackage{authblk} % for authors
\usepackage{xcolor}
\usepackage{physics}
\usepackage{graphicx} % for graphics in pdf_tex figures
\usepackage{float}
\usepackage{tikz}
\usetikzlibrary{calc}
\usepackage{comment}

\usepackage{acronym}
\newacro{EFT}[EFT]{effective field theory}
\acrodefplural{EFT}{effective field theories}
\newacro{UV}[UV]{ultraviolet}
\newacro{IR}[IR]{infrared}
\newacro{AdS}[AdS]{anti-de Sitter}
\newacro{BF}[BF]{Breitenlohner-Freedman}

\usepackage{hyperref}
\usepackage{varioref}       
\usepackage{cleveref}     

\crefname{table}{table}{tables}
\Crefname{table}{Table}{Tables}
\crefname{figure}{figure}{figures}
\Crefname{figure}{Figure}{Figures}

\definecolor{tealblue}{rgb}{0.21, 0.56, 0.63}
\hypersetup{colorlinks=true,allcolors = tealblue,linktocpage=true}

\usepackage{pgf} % for pgf pics, might have to remove on publication if pgfs have to be prerendered

\usepackage[
        backend=bibtex,
        sorting=none,
        style=phys,
        eprint=true,
        doi=false,
        biblabel=brackets
        ]{biblatex}

\bibliography{casimir}

\usepackage[backgroundcolor=white!95!red, linecolor=cyan, bordercolor=cyan, textsize=footnotesize]{todonotes}

\newcommand{\commie}[1]{}

\numberwithin{equation}{section}
\numberwithin{table}{section}

\newenvironment{eqaed}
    {\begin{equation}
    \begin{aligned}
    }
    { 
    \end{aligned}
    \end{equation}
    \ignorespacesafterend
    }

\begin{document}

\title{Instabilities in scale-separated Casimir vacua}
\date{}

\author{Miquel Aparici\thanks{maparici@mpp.mpg.de} }

\author{Ivano Basile\thanks{ibasile@mpp.mpg.de}}
\affil{\emph{Max-Planck-Institut f\"ur Physik (Werner-Heisenberg-Institut)}\\ 
\emph{Boltzmannstraße 8, 85748 Garching, Germanythen}}

\author{Nicolò Risso\thanks{nicolo.risso@studenti.unipd.it}}
\affil{\emph{ Dipartimento di Fisica e Astronomia “Galileo Galilei”, }\\ \emph{Università degli Studi di Padova,
and I.N.F.N. Sezione di Padova}\\ \emph{ Via F. Marzolo 8, 35131 Padova, Italy}}

\maketitle

\begin{abstract}
    
    \noindent Parametric scale separation is notoriously difficult to achieve in flux compactifications of gravitational effective theories. An appealing alternative to conventional Freund-Rubin vacua involves Ricci-flat internal manifolds, where the energy supplied by fluxes is balanced not by curvature but by the Casimir energy. The internal volume can be stabilized by this mechanism producing anti-de Sitter geometries with parametric scale separation, including an explicit example in eleven-dimensional supergravity. We study deformations of these geometries, showing the presence of perturbative and non-perturbative instabilities.
    
\end{abstract}

\thispagestyle{empty}

\newpage

\tableofcontents

\thispagestyle{empty}

\newpage

\pagenumbering{arabic}

\section{Introduction}\label{sec:introduction}

Among the several requirements for a model of cosmology and particle physics to be realistic, perhaps one of the most basic ones is to reproduce genuine four-dimensional physics for low-energy observers. Broadly speaking, we have not observed any phenomena that ultimately cannot be accounted by gravitational \ac{EFT} in four dimensions. The energy scale at which new physics appears is bounded by the Planck scale, but it could be much lower --- in particular, in the presence of large extra dimensions, which are common in string constructions, physics would qualitatively change above the Kaluza-Klein gap. Since any such physics is separated from the scales probed by high-energy experiments, it is also separated from the Hubble parameter which sets the effective scale of the observable universe. This is the requirement of \emph{scale separation}. In the context of extra dimensions, where the scale of new physics is set by the Kaluza-Klein gap, an intuitive phrasing of this condition is that the internal manifold be much smaller than the observable universe. This is the typical intension used in the literature \cite{Coudarchet:2023mfs}, although a realistic model need \emph{a priori} only satisfy the weaker, more general notion of scale separation we outlined.

In an \ac{UV}-complete context, geometric compactifications are in principle a restriction, since non-geometric settings without extra dimensions exist in string theory \cite{ Kawai:1986va, Narain:1986qm, Lerche:1986cx, Antoniadis:1986rn, Dixon:1987yp,  Candelas:1987kf, Antoniadis:1987wp, Gepner:1987qi, Green:1988wa, Green:1988bp, Candelas:1988di, Kazama:1988qp, Vafa:1988uu, Greene:1988ut, Candelas:1989ug, Candelas:1989js, Narain:1990mw, Witten:1993yc, Kachru:1995wm, Angelantonj:1996mw, Banks:1996vh, Ishibashi:1996xs, Dijkgraaf:1997vv, Blumenhagen:1998tj, Bianchi:1999uq, Israel:2013wwa, Hull:2017llx, Gkountoumis:2023fym, Baykara:2023plc, Becker:2024ayh, Rajaguru:2024emw, Chen:2025rkb}. However, whenever the \ac{EFT} cutoff is parametrically smaller than the Planck scale there is substantial evidence that mesoscopic extra dimensions are present \cite{Aoufia:2024awo, Ooguri:2024ofs}. The only alternative seems to be that new physics is driven purely by weakly coupled strings \cite{Lee:2018urn, Lee:2019wij, Lee:2019xtm, Stout:2021ubb, Stout:2022phm, Basile:2023blg, Bedroya:2024ubj, Kaufmann:2024gqo, Herraez:2024kux, Herraez:2025clp, Herraez:2025gcf, Friedrich:2025gvs}. However, in that case, the extra degrees of freedom required to obtain only four extended spacetime dimensions are also gapped at the string scale. Either way, the goal is to realize a genuinely four-dimensional \ac{EFT} without additional long-range forces\footnote{Moreover, in the presence of moduli, low-energy observers can have access to the extra dimensions when probing black holes \cite{Sen:2025ljz, Sen:2025oeq, Castellano:2025ljk, Calderon-Infante:2025pls, Delgado:2025crl, Castellano:2025yur, Castellano:2025rvn}.} mediated by moduli. The latter issue is the problem of \emph{moduli stabilization}, and it is intimately connected to the story of scale separation, since it is straightforward to realize scale-separated (Minkowski) compactifications without stabilizing moduli.

This paper focuses on the simplest set of ingredients which lead to scale-separated \acp{EFT}, namely geometric flux compactifications over flat manifolds supported by Casimir energy. These are well-understood settings which arise naturally in higher-dimensional supergravity, where the full contribution to vacuum energy, when supersymmetry is softly broken \`{a} la Scherk-Schwarz, is given by the \ac{UV}-insensitive Casimir term. Furthermore, these theories naturally contain fluxes and sometimes are known to arise in the string landscape. The most prominent example --- and, as we shall see, essentially the only one meeting all our requirements --- is eleven-dimensional supergravity, the low-energy limit of M-theory (in its maximal number of extended dimensions). It was shown in \cite{Luca:2022inb} that the four-form flux of eleven-dimensional supergravity can combine with the Casimir energy of the massless fields on a seven-torus, yielding a simple \ac{AdS} solution in four large dimensions. This solution enjoys parametric scale separation in terms of the flux quantum $N \gg 1$, which also suppresses higher-derivative corrections.

The analysis of \cite{Luca:2022inb} shows that the volume mode of the internal torus lies at a minimum of the effective potential. However, the ansatz of a square torus need not be preserved by perturbations, which can be unstable. Since supersymmetry if broken via Scherk-Schwarz twists, \emph{a priori} one should also consider deformations of the torus which produce curvature. Since the Casimir contribution to the effective potential is computed via field fluctuations around the background, the solution and the stability analysis have to be performed self-consistently around the deformed geometry. The main aim of this paper is to build the mathematical toolkit to do so. Using a combination of heat-kernel and Green-function methods, we shall compute the effective potential for various types of deformations of the internal torus of \cite{Luca:2022inb}, including ones that produce curvature. We will prove explicitly that none of them contribute tadpoles to the effective field equations. However, we identified a tachyonic direction in the space of flat volume-preserving deformations, whose squared mass is negative and lies well below the BF bound for perturbative stability. At any rate, as we will argue, any putative classically stable vacuum would undergo non-perturbative decay via brane nucleation; however, this decay channel is far less severe, and may be physically acceptable for the purposes of deriving weakly coupled, genuinely lower-dimensional \acp{EFT}.

The content of the paper are organized as follows. In \cref{sec:casimir_vacua} we review the solution of \cite{Luca:2022inb} in detail, generalizing it to other dimensions and discussing other potentially relevant cases. In \cref{sec:casimir_torus} we derive useful formulae to compute the Casimir energy for tori and their deformations. The analysis is somewhat complementary to the more systematic and exhaustive framework provided in \cite{ValeixoBento:2025yhz, DallAgata:2025jii}, since we also discuss curvature deformations in \cref{sec:casimir_torus_pert}. Armed with this toolkit, in \cref{sec:tadpoles} we prove that the solution is devoid of tadpoles in any direction in the space of internal perturbations. In \cref{sec:tachyons} we assess the stability of the solution: in particular, in \cref{sec:stability} we show the presence of a \ac{BF} tachyon, while in \cref{sec:general_decay} we argue that M2-brane nucleation is a universally allowed decay channel even in the absence of \ac{BF} tachyons. We provide some closing remarks for future investigations in \cref{sec:conclusions}.

\section{Candidate vacua with parametric scale separation}\label{sec:casimir_vacua}

In the following, we review the construction of semiclassical flux compactifications supported by Casimir energy. Thanks to this contribution it is possible to obtain solutions with flat internal spaces and achieve parametric scale separation in the weakly curved regime. The simplest setting is M-theory, whose low-energy description in eleven dimensions contains no additional bosonic fields to account for.

In \cite{Luca:2022inb}, a novel construction of non-supersymmetric scale-separated \ac{AdS} solutions of eleven-dimensional supergravity was presented. The Casimir energy is used as ``negative energy'' source to violate the Reduced Energy Condition, in order to bypass no-go theorems against scale separated \ac{AdS} vacua \cite{Gautason:2015tig, DeLuca:2021mcj}. Here we will review briefly the details of the solution and their generalizations.

We begin from M-theory in its low-energy description in terms of eleven-dimensional supergravity, and consider an \ac{AdS}$_4$ compactification on a seven-dimensional square torus. The ansatz for the metric is
\begin{eqaed}
    ds^2_{11} = L^2 ds^2_{\text{AdS}_4} + R^2 ds^2_{T^7} \,,
\end{eqaed}
where the metrics on the \ac{AdS}$_4$ and $T^7$ factors have been chosen to have unit radii of curvature. Before introducing the Casimir energy, the bosonic effective action takes the form
\begin{eqaed}
    S_\text{eff} = \frac{1}{2\kappa^2} \int d^{11}x \sqrt{-g} \left(\mathcal{R} - \frac{1}{2 \cdot 4!} \, \abs{F_4}^2 \right) ,
\end{eqaed}
where the gravitational constant is $2\kappa^2 \equiv (2\pi)^8 \ell_{\text{Pl},11}^9 = 2\pi M_{\text{Pl},11}^{-9}$ in terms of the eleven-dimensional Planck length and mass. We denote the Ricci scalar by $\mathcal{R}$ in order to avoid confusion with the internal radius $R$. 

The next step is including the Casimir energy of the massless fields, namely the metric $g$, the four-form $F_4$ and the gravitino $\psi$, since the contribution from massive fields is exponentially suppressed. In order to obtain a non-trivial result, supersymmetry is broken by imposing anti-periodic boundary conditions for fermions on the torus cycles. The full computation can be performed explicitly for the flat background; as we shall show in \cref{sec:AdS_casimir}, this approximation is valid \emph{a posteriori}. However, for the purposes of finding the solution it is enough to specify the sign of the Casimir energy and its scaling with $R$. The latter follows on dimensional grounds, while the overall sign is negative, as contributed by the bosons, due to the Scherk-Schwarz twist. These requirements bring us to the effective contribution\footnote{In our conventions, the factor of two in \cref{eq:casimir_action} matches the solutions of \cite{Luca:2022inb} including the numerical prefactors.}
\begin{eqaed}\label{eq:casimir_action}
    S_\text{Casimir} = \frac{2|\rho_c|}{(2\pi)^8}  \int_{M_{11}} d^{11}x \, \sqrt{-g} \, R^{-11} \,,
\end{eqaed}
where $|\rho_c|$ is a numerical factor depending on the topology and number of fields that we will compute explicitly, but does not affect the ensuing considerations. The resulting stress-energy tensor does not satisfy the Reduced Energy Condition \cite{Luca:2022inb}.

To stabilize the size of the torus, we need a positive energy contribution from the flux of $F_4$, whose profile is taken to be
\begin{eqaed}
    F_4 = f_4 \operatorname{vol}_{\text{AdS}_4} \,,
\end{eqaed}
with $f_4$ a real constant. Flux quantization of the magnetic dual (which in our conventions is given by $F_7 := \star F_4 = -f_4 \frac{R^7}{L^4}\operatorname{vol}_{T^7}$) yields
\begin{eqaed}\label{eq:quantization_condition}
    \frac{1}{(2\pi \ell_{\text{Pl},11})^6} \int_{T^7} F_7 = N \quad \Longrightarrow \quad f_4^2 = \frac{N^2}{4\pi^2}\ell_{\text{Pl},11}^{12}\frac{L^8}{R^{14}} \,.
\end{eqaed}
Solving the field equations with these sources and ansätze, one finds
\begin{eqaed}
    L^2 = \ell_{\text{Pl},11}^2 \left ( \frac{N}{2\pi}\right)^{\frac{22}{3}} |\rho_c|^{-\frac{14}{3}} \frac{7^{14/3}}{2^{11} \cross 3^{8/3}}   \, , \quad R^{11} = \ell_{\text{Pl},11}^{11} \left ( \frac{N}{2\pi}\right)^{\frac{22}{3}}  |\rho_c|^{-\frac{14}{3}} \frac{7^{11/3}}{2^{11} \cross 3^{11/3}} \,.
\end{eqaed}
In the large-$N$ limit, both radii are large in Planck units, providing parametric control over corrections. In addition, in this limit the solution exhibits scale separation, since
\begin{eqaed}
    \frac{R^2}{L^2} \propto N^{-6} \,.
\end{eqaed}
This is an example of a more general class of $\text{AdS}_d \times T^q$ solutions with $q>d$ and $q$-dimensional magnetic fluxes $\int_{T^q} \star F_{d-q}$, found by extremizing a reduced effective action $S_\text{eff} \equiv \frac{1}{\kappa^2} \int S_\text{red}$ given by
\begin{eqaed}\label{eq:redS}
    S_\text{red} = L^d R^q \left (-\frac{d(d-1)}{2L^2} - \frac{\ell_{\text{Pl},d+q}^{2q-2}}{4R^{2q}}\left ( \frac{N}{2\pi} \right )^2 + \frac{|\rho_c|\ell_{\text{Pl},d+q}^{d+q-2}}{R^{d+q}} \right ) ,
\end{eqaed}
where the first term comes from the AdS scalar curvature, the second from the kinetic term of fluxes and the third one from the Casimir energy. When solutions that extremize this effective action exist, they have radii given by
\begin{eqaed}\label{eq:radii}
    R & = \ell_{\text{Pl},d+q} \left ( \frac{N}{2\pi}\right )^{\frac{2}{q-d}} |\rho_c|^{-\frac{1}{q-d}} \left ( \frac{q(d-1)}{2d(d+q-2)}  \right )^{\frac{1}{q-d}} \, , \\
    L & = \ell_{\text{Pl},d+q} \left ( \frac{N}{2\pi}\right )^{\frac{q+d}{q-d}} |\rho_c|^{-\frac{q}{q-d}} \, \sqrt{\frac{d-1}{q-d}} \left( 2d(d+q-2)\right)^{-\frac{d+q}{2(q-d)}} \left(q(d-1)\right)^{\frac{q}{q-d}} \, .
\end{eqaed}
Hence, they exhibit parametric scale separation, since 
\begin{eqaed}
    \frac{R}{L } \sim N^{\frac{2-d-q}{q-d}} \ll 1 \, .
\end{eqaed}
Plugging this class of backgrounds into a generic higher-derivative contribution to the effective action, we obtain a sub-leading result with respect to the leading two-derivative action in the large-$N$ limit, schematically
\begin{eqaed}
    \frac{\mathcal{R}^n \abs{F_4}^m}{\mathcal{R}} \sim \frac{\mathcal{R}^n \abs{F_4}^m}{\abs{F_4}^2} \sim N^{-(2n+m-2)\frac{d+q}{q-d}} \ll 1 \, .
\end{eqaed}
This points to the higher-derivative corrections to these solutions being parametrically under control. Generally speaking, at large volume there can still be small cycles supporting light states beyond the \ac{EFT} regime. If these states come from wrapped M-branes, they could significantly affect our analysis for sub-Planckian cycles. Here we study a neighborhood of the square-torus point in the classical moduli space, which is safely away from these regions; similarly, other extrema obtained from the Casimir energy should satisfy this condition.

The results above hold whenever $q>d>1$, although this does not \emph{a priori} address whether such solutions can be actually realized in the string landscape. It would be interesting to carry out a more detailed study of such examples including whether they can be embedded in string theory. This is beyond the scope of this paper, but we can collect a few remarks. The Casimir energy itself is not \ac{UV}-sensitive, but the gravitational field should couple to the complete vacuum energy. Its \ac{UV}-sensitive part vanishes in supergravity, so that no \ac{UV}-completion is needed to obtain a reliable solution. This is to be contrasted with \emph{e.g.} the heterotic solutions of \cite{Mourad:2016xbk, Basile:2018irz, Antonelli:2019nar}, where the full one-loop vacuum energy provides the dilaton potential. Since the theories we consider do not contain scalar fields, they exclude the usual perturbative string settings in which the full vacuum energy can be computed. Thus, absent fine-tunings or mechanisms akin to Atkin-Lehner symmetry \cite{Moore:1987ue, Dienes:1990qh, Gannon:1992su}, it is natural to ask which supergravities would lead to such solutions. Besides eleven-dimensional supergravity, the only candidate is pure supergravity with eight supercharges in five dimensions. This would provide \ac{AdS}$_2$ vacua. Even if perturbatively stable, these solutions would lead to less interesting lower-dimensional \acp{EFT} without gravitons. Nevertheless, if no perturbatively stable Casimir vacua were to exist in eleven-dimensional supergravity, these settings would be a natural place to look next, although no stringy realization of minimal five-dimensional supergravity is known \cite{Baykara:2023plc}.

In the remainder of the paper we build the necessary tools to assess the stability of the proposal in \cite{Luca:2022inb}. The explicit form of the Einstein frame 4-dimensional effective potential for the radion, as we will see, shows a minimum in the volume direction. However, this leaves open the possibility of tachyonic modes in the spectrum or even of entirely off-shell directions. As we shall discuss in detail, it turns out that the solution is on-shell but perturbatively (and non-perturbatively) unstable. The framework we will now present can be readily applied to more complicated settings, such as those discussed in \cite{ValeixoBento:2025yhz, DallAgata:2025jii}.

\section{Casimir energy for flat tori}\label{sec:casimir_torus}
\subsection{Warm-up: Casimir energy on a circle}\label{sec:casimir_circle}

As a warm-up and overview of various methods to compute Casimir energies, let us consider a circle compactification $\mathbb{R}^{d-1,1} \times S^1_R$, where the circle has radius $R$. We will only consider massless bosonic fields, since the fermionic contributions will not cancel the bosonic contributions due to the Scherk-Schwarz twist. A real massless scalar field decomposes in Kaluza-Klein modes with masses $m_n = \frac{n}{R}$. A straightforward way to write down the Casimir contribution to the vacuum energy (density) is the ``trace-log'' expression of the (Wick-rotated) one-loop effective action. Subtracting the corresponding contribution in $\mathbb{R}^{d,1}$, one finds
\begin{eqaed}\label{eq:eff_pot_circle}
    V = \frac{1}{2} \int \frac{d^dp}{(2\pi)^d} \left( \frac{1}{2\pi R}\sum_{n \in \mathbb{Z}} \log \left(p^2 + \frac{n^2}{R^2}\right) - \, \int \frac{dk}{2\pi} \log \left(p^2 + k^2\right) \right) \, , 
\end{eqaed}
where a regulator is implied. The limit in which the regulator is removed exists and is finite, and yields the physical Casimir energy, with corresponding force (density) $F = - \frac{dV}{dR}$. The above expression can be recast in a manifestly finite form, which will be useful when discussing deformations. This can be achieved via the heat kernel expression of the one-loop effective action, which can be obtained via the Schwinger proper-time integral
\begin{eqaed}\label{eq:eff_pot_circle_schwinger}
    V & = - \, \frac{1}{2} \int \frac{d^dp}{(2\pi)^d} \int_0^\infty \frac{ds}{s} \left(\frac{1}{2\pi R}\sum_{n \in \mathbb{Z}} \, e^{-\left(p^2 + \frac{n^2}{R^2}\right)s} - \, \int \frac{dk}{2\pi} \, e^{-\left(p^2 + k^2\right)s} \right) \\
    & = - \, \frac{\pi^{\frac{d}{2}}}{2(2\pi R)^{d+1}} \int_0^\infty \frac{ds}{s^{\frac{d}{2}+1}} \left( \theta_3(e^{-s}) - \, \sqrt{\frac{\pi}{s}} \right) ,
\end{eqaed}
where $\theta_3(e^{-s}) = \sum_{n \in \mathbb{Z}} e^{-n^2 s}$ is the third Jacobi theta function. In order to extract the value of this integral one can perform a Poisson resummation, but since we will deal with non-integrable deformations it is more instructive to use another method to find the result for the unperturbed case \cite{Arkani-Hamed:2007ryu}. One can compute the energy density $V = \langle T_{00} \rangle$ from the expectation value of the energy-momentum tensor. For a free scalar field, one finds
\begin{eqaed}
    \langle T_{00} \rangle = \lim_{x \to x'} \frac{\partial}{\partial x^0} \frac{\partial}{\partial {x'}^0} \, G(x,x') \, ,
\end{eqaed}
where $G$ denotes the propagator. On the background $\mathbb{R}^{d-1,1} \times S^1_R$, where the coordinate along the circle is $y$, the method of images yields\footnote{The result in \cref{eq:circle_propagator} only holds for $d>1$. For $d=1$ the two-dimensional propagator is logarithmic, but the final result for the Casimir energy in \cref{eq:casimir_circle} is still valid.}
\begin{eqaed}\label{eq:circle_propagator}
    G(x,x') = \frac{1}{(d-1)\Omega_d} \, \sum_{n \in \mathbb{Z}} \frac{1}{\abs{x-x' + 2\pi R \, n \, \hat{e}_y}^{d-1}} \, ,
\end{eqaed}
where $\Omega_d \equiv \frac{2\pi^{\frac{d+1}{2}}}{\Gamma\left(\frac{d+1}{2}\right)}$ denotes the volume of the unit $d$-sphere. Thus, subtracting the $\mathbb{R}^{d,1}$ contribution and taking into account the Wick rotation when differentiating with respect to Euclidean time, one finds
\begin{eqaed}\label{eq:casimir_circle}
    V_{S^1}(R) = - \, \frac{1}{\Omega_d} \, \sum_{n \neq 0} \frac{1}{\abs{2\pi R \, n \, \hat{e}_y}^{d+1}} = - \, \frac{2\zeta(d+1)}{\Omega_d (2\pi R)^{d+1}} \, ,
\end{eqaed}
where $\zeta$ is the Riemann zeta function. For other free bosonic fields, the result is multiplied by the number of physical polarizations.

So far, we have computed the Casimir energy for bosonic species. For fermionic fields, the one-loop effective potential can vary depending on the spin structure chosen for the internal manifold. For a single circle $S^1$, there are two possible spin structures: one corresponding to periodic boundary conditions and the other to antiperiodic boundary conditions for spinors. For periodic boundary conditions, the result is the same as for bosonic fields, except that an additional negative sign appears in front of the Casimir potential due to fermion number. In the case of antiperiodic boundary conditions, the Kaluza-Klein modes acquire masses $m_n = \frac{n + \frac{1}{2}}{R}$, and \cref{eq:eff_pot_circle} still holds, up to summing over half-integer Kaluza-Klein modes. In addition, an overall minus sign appears in front due to fermion number. As a result, in the Schwinger integral of \cref{eq:eff_pot_circle_schwinger} the second Jacobi theta function $\theta_2(e^{-s})=\sum_{n\in \frac{1}{2}\mathbb{Z}} e^{-n^2s}$ will appear instead of $\theta_3(e^{-s})$. The Green function (\ref{eq:circle_propagator}) is similarly modified. The non-supersymmetric spin structure is accounted for by adding an oscillating factor $(-1)^n$ in the general term of the sum. All in all, taking into account the extra negative sign coming from fermion number, the Casimir energy reads
\begin{eqaed}\label{eq:casimir_circle_fermion}
    V_{S^1}(R) = \, \frac{1}{\Omega_{d}} \, \sum_{n \neq 0} \frac{(-1)^{n}}{\abs{2\pi R \, n}^{d+1}} \, .
\end{eqaed}
This expression can be rewritten as a multiple of the ordinary zeta function according to
\begin{eqaed}\label{eq:twisted_zeta}
    \sum_{n \neq 0} \frac{(-1)^n}{\abs{n}^s} = (2^{1-s}-1) \, 2 \zeta(s) \, ,
\end{eqaed}
showing that fermionic contributions do not cancel bosonic contributions for Scherk-Schwarz compactifications.

Lastly, consider a theory in which there are the same number of bosonic and fermionic species. If the periodic spin structure is chosen for the fermions, the contributions from the fermionic and bosonic fields are identical but with opposite signs, leading to the typical supersymmetric cancellation. In contrast, for antiperiodic boundary conditions, the contributions differ, resulting in a non-zero Casimir energy. Interestingly, in this case, ultraviolet divergences from the bosons and fermions still cancel identically, since they are local effects corresponding to flat spacetime. Thus, the result is finite without subtracting the contribution of the uncompactified background.

\subsection{Casimir energy on a flat torus}

The above method can be straightforwardly generalized to compactifications on square tori $\mathbb{R}^{d-1,1} \times T^q_R$, where all radii are equal to $R$. \Cref{eq:circle_propagator} generalizes to
\begin{eqaed}\label{eq:torus_propagator}
    G(x,x') = \frac{1}{(d+q-2)\Omega_{d+q-1}} \, \sum_{n \in \mathbb{Z}^q} \frac{1}{\abs{x-x' + 2\pi R \, \vec{n}}^{d+q-2}} \, ,
\end{eqaed}
where $\mathbf{n}$ denotes the vector with vanishing components in spacetime and internal components $n \in \mathbb{Z}^q$. As a result, the Casimir energy (density) becomes
\begin{eqaed}\label{eq:casimir_torus}
    V_{T^q}(R) = - \, \frac{1}{\Omega_{d+q-1}} \, \sum_{n \neq 0} \frac{1}{\abs{2\pi R \, \vec{n}}^{d+q}} = - \, \frac{\zeta_{\mathbb{Z}^q}(d+q)}{\Omega_{d+q-1} (2\pi R)^{d+q}} \, ,
\end{eqaed}
where now $\zeta_{\Lambda}(s) \equiv \sum_{n \in \Lambda - \{0\}} \frac{1}{\abs{n}^s}$ denotes the Epstein zeta function. For fields with antiperiodic boundary conditions along the torus circles the result is the same, but there is an oscillating factor $(-1)^{\sum_i n_i}$ in the general term of the sum. These sums converge rather rapidly and can be estimated numerically. This method is useful to obtain the numerical value of the energy density for square tori, which can be used as a benchmark for other methods. However, since it hinges on the position-space representation of the propagator, it is difficult to apply directly to deformed geometries. We will thus use this result to compare with other approaches based on the heat kernel and propagator.

These results are valid for an external Minkowski spacetime. As we shall show shortly, they are valid for scale-separated AdS as well. In order to see this more directly, it is useful to review other methods of computing the Casimir energy. These methods will allow us to include internal deformations as well. 

\subsection{Heat kernel method}\label{sec:heat_kernel}

As we have anticipated, for our purposes it is instructive to recast the above result in a Schwinger-like form using the heat kernel. In order to extend the computations to deformed geometries, let us first use a schematic notation before specializing to the case of a square torus. The heat kernel $K(x,y \,|\, s)$ associated to a kinetic differential operator $L$ acting on the Hilbert space of square-integrable functions (with respect to the $x$ coordinate) solves
\begin{eqaed}\label{eq:heat_kernel_def}
    \left(\partial_s + L\right) K = 0
\end{eqaed}
subject to the distributional initial condition $K \overset{s \to 0^+}{\longrightarrow} \delta(x,y)$. Labeling an orthogonal spectral basis\footnote{More precisely, in the cases we are going to consider, the $f_i(x)$ are actually non-normalizable plane waves, to be treated with the standard rigged space construction.} $\{f_i(x)\}_i$ of $L$ with a (possibly continuous) index $i$ and the respective eigenvalues with $\lambda_i$, one has
\begin{eqaed}\label{eq:heat_kernel_gen}
    K = \sum_i \, f_i(x)^* \, f_i(y) \, e^{- \lambda_i s} \, .
\end{eqaed}
The effective (Euclidean) Lagrangian density then reads \cite{Vassilevich:2003xt}
\begin{eqaed}\label{eq:heat_kernel_eff_action}
    \mathcal{L}_\text{eff}^\text{E} = - \, \frac{1}{2} \int_0^\infty \frac{ds}{s} \, K(x,x \, | \, s) \, .
\end{eqaed}
Importantly, the heat kernel is multiplicative on product Riemannian manifolds. For flat spacetime $\mathbb{R}^{d+q-1,1}$ one has
\begin{eqaed}\label{eq:heat_kernel_flat}
    K(x,y \, | \, s) = \int \frac{d^{d+q}p}{(2\pi)^{d+q}} \, e^{-p^2 s + i p \cdot (x-y)} \quad \longrightarrow \quad K(x , x \, | \, s) = \frac{1}{(2\pi)^{d+q}} \, \left(\frac{\pi}{s}\right)^{\frac{d+q}{2}} \, ,
\end{eqaed}
while for the background $\mathbb{R}^{d-1,1} \times T^q_R$
\begin{eqaed}\label{eq:heat_kernel_torus}
    K = \int \frac{d^{d}p}{(2\pi)^{d}} \sum_{n \in \mathbb{Z}^q} \frac{e^{-\left(p^2+\frac{n^2}{R^2}\right) s + i \left(p+\frac{\mathbf{n}}{R}\right) \cdot (x-y)}}{(2\pi R)^q} \longrightarrow K(x , x \, | \, s) = \frac{\theta_3(e^{-\frac{s}{R^2}})^q}{(2\pi)^{d+q}R^q} \, \left(\frac{\pi}{s}\right)^{\frac{d}{2}} .
\end{eqaed}
Putting things together, the Casimir energy density for square tori then reads
\begin{eqaed}\label{eq:casimir_torus_heat_kernel}
    V = - \, \frac{1}{2(2\pi R)^{d+q}} \int_0^\infty \frac{ds}{s} \left(\frac{\pi}{s}\right)^{\frac{d}{2}} \left( \theta_3(e^{-s})^q - \left(\frac{\pi}{s}\right)^{\frac{q}{2}} \, \right) \, .
\end{eqaed}
Once again, one may recast the result in terms of zeta functions with a Poisson resummation. Upon deforming the torus, the heat kernel will turn out to be more complicated than the propagator, but it will be useful to keep the above integral expression in mind. Indeed, the final result obtained from the perturbed propagator can be recast in a similar form, which is manifestly finite without needing to introduce a regulator. This is because of the crucial property
\begin{eqaed}\label{eq:theta_asymptotics}
    \theta_3(e^{-s}) \overset{s \to 0^+}{\sim} \sqrt{\frac{\pi}{s}} \left( 1 + 2 \, e^{-\frac{\pi^2}{s}} \right) \, ,
\end{eqaed}
which will prove useful to show finiteness of the analog of \cref{eq:casimir_torus_heat_kernel} for the deformed geometry that we will consider.

\subsubsection{Casimir energy in scale-separated AdS}\label{sec:AdS_casimir}

The approach we outlined above has the advantage of showing directly why the flat background is a valid approximation for the $\text{AdS}_d \times T^q$, insofar as the curvature radius $L$ of AdS is parametrically larger than the radius $R$ of the torus, here still taken to be square for simplicity. The Casimir energy in this case takes the form
\begin{eqaed}\label{eq:AdS_casimir}
    V_\text{AdS} = - \, \frac{1}{2(2\pi)^{q}} \int_0^{\infty} \frac{ds}{s} \, K_\text{AdS}(s) \left( \frac{\theta_3(e^{-\frac{s}{R^2}})^q}{R^q} - \left(\frac{\pi}{s}\right)^{\frac{q}{2}} \, \right) \, .
\end{eqaed}
Rescaling $s$ as in \cref{eq:casimir_torus_heat_kernel} recasts the integrand into a function of the dimensionless argument $\frac{R^2}{L^2}s$, and thus the integral into a function of $\frac{R^2}{L^2} \ll 1$. For generic $s$, the local heat-kernel expansion of the AdS factor \cite{Vassilevich:2003xt} is dominated by the flat-spacetime expression, which is corrected by powers of the scale-separation parameter. However, exchanging this expansion with the Schwinger integral produces infrared divergences, since the non-local corrections to the effective action from integrating out a massless field cannot be captured by the local version of the heat-kernel expansion. Put differently, for massive fields the above procedure would work, although the limit $R \ll L$ and the massless limit of this procedure are affected by an ordering issue.

Nevertheless, this incomplete argument for the validity of the flat-background approximation to the Casimir energy leads to the correct conclusion. To see this in the case of interest of massless fields, we can use the explicit form of the heat kernel for Euclidean AdS, namely hyperbolic space \cite{10.1112/S0024609398004780}. Let $\rho = d(x,y)$ be the hyperbolic distance between two points $x,\,y$. For an odd number of dimensions $d=2m+1$, the heat kernel in hyperbolic space reads 
\begin{equation}
    K_{\text{AdS}_d}(x,y\,|\,s) = \frac{(-1)^m}{(2\pi)^m}\frac{1}{(4\pi s)^{\frac{1}{2}}L^{m}} \left (\frac{1}{\sinh{\frac{\rho}{L}}} \frac{\partial}{\partial\rho}\right )^m e^{-\frac{m^2s}{L^2}-\frac{\rho^2}{4s}} \, ,
\end{equation}
whereas for even dimension $d=2m$ we have
\begin{equation}
    K_{\text{AdS}_d}(x,y\,|\,s) = \frac{\sqrt{2}(-1)^m e^{-\frac{(2m+1)^2}{4}\frac{s}{L^2}}}{(2\pi)^m (4\pi s)^{3/2}L^{m+1}} \left (\frac{1}{\sinh{\frac{\rho}{L}}} \frac{\partial}{\partial\rho}\right )^m \hspace{-8pt} \int_{\rho}^{\infty}\hspace{-8pt}\frac{u e^{-\frac{u^2}{4s}} du}{\sqrt{\cosh \left (\frac{u}{L}\right )-\cosh\left ( \frac{\rho}{L}\right)}} \, .
\end{equation}
In the odd-dimensional case, the expression for the heat kernel is sufficiently simple that we can directly argue the validity of the flat-space approximation. In general, after taking the coincidence limit, the heat kernel has the form
\begin{equation}
    K_{\text{AdS}_d}(s) = K_{\mathbb{R}^d}(s) e^{-\frac{(d-1)^2}{4}\frac{s}{L^2}}\left(1+\sum_{n=1}^{d-2} a_n \left(\frac{s}{L^2}\right)^n\right ) ,
\end{equation}
where $K_{\mathbb{R}^d}(s)=\frac{1}{(4\pi s)^{d/2}}$ is the heat kernel of flat space. For instance,
\begin{eqaed}
    K_{\text{AdS}_3}(s) = \frac{e^{-\frac{s}{L^2}}}{(4\pi s)^{\frac{3}{2}}}\,, \quad K_{\text{AdS}_5}(s) = \frac{e^{-\frac{4s}{L^2}}}{(4 \pi s)^{\frac{5}{2}}}\left (1+\frac{2}{3}\frac{s}{L^2}\right )\,, \\ K_{\text{AdS}_7}(s) = \frac{e^{-\frac{4s}{L^2}}}{(4 \pi t)^{\frac{7}{2}}}\left (1+2\frac{s}{L^2}+\frac{16}{15}\frac{s^2}{L^4}\right )\,.
\end{eqaed}
After performing the Schwinger integral, each of the additional terms with respect to the flat space heat kernel will scale as a power of the quotient of the radii $R/L$. Thus, they vanish in the scale separation limit. To see how the flat space result is recovered as $R/L$ tends to zero, we perform a Poisson resummation in (\ref{eq:AdS_casimir}). Defining $r := \frac{d-1}{2} \frac{R}{L}$, we write
\begin{eqaed}
    V_\text{AdS} &= - \frac{\pi^{\frac{d+q}{2}}}{2(2\pi R)^{q+d}} \sum_{\vec{k}\neq0} \int_0^{\infty} \frac{ds}{s^{\frac{q+d}{2}+1}} e^{-r^2s} e^{-\frac{\vec{k}^2\pi^2}{s}} + \mathcal{O}\left(\frac{R^2}{L^2}\right) \\
    &= - \frac{1}{(2\pi R)^{q+d}} \sum_{\vec{k}\neq 0} \frac{1}{|\vec{k}|^{d+q}} \left( r|\vec{k}| \right)^{\frac{d+q}{2}} K_{\frac{d+q}{2}}( 2\pi r |\vec{k}|) +\mathcal{O}\left(\frac{R^2}{L^2}\right).
\end{eqaed}
Here, $K_n(x)$ is the modified Bessel function of the second kind. Using that
\begin{equation}
    \lim_{x\rightarrow0^+}x^n K_n(2\pi x) = \frac{1}{\Omega_{2n-1}} \, ,
\end{equation}
we see that the flat space result (\ref{eq:casimir_torus}) is recovered in the limit $R/L\rightarrow0$. If the dimension of the AdS is even there is not an analogous argument. Instead, we have numerically computed the integral (\ref{eq:AdS_casimir}) for $d = 2, 4$ and confirmed that the Casimir potential indeed approaches the flat-space result as $R/L$ tends to zero.

\subsection{Green function method}\label{sec:propagator}

For the purposes of deforming the geometry of the torus, it will prove more convenient to use the propagator. In the general schematic notation used above, it takes the form
\begin{eqaed}\label{eq:heat_kernel_propagator}
    G(x,y) = - \, \int_0^\infty ds \, K(x, y \, | \, s) = - \, \sum_i \frac{f_i(x)^* \, f_i(y)}{\lambda_i} \, ,
\end{eqaed}
which satisfies $L \, G = - \, \delta$. Computing the energy density from the energy-momentum tensor in the background $\mathbb{R}^{d-1,1} \times T^q_R$ gives
\begin{eqaed}\label{eq:casimir_torus_propagator_EM_tensor}
    V = - \, \frac{1}{(2\pi R)^{d+q}} \int d^dp \, p_0^2 \left(\sum_{n \in \mathbb{Z}^q} \frac{1}{p^2 + n^2} - \int_{\mathbb{R}^q} \frac{d^qn}{p^2 + n^2} \right) \, ,
\end{eqaed}
where the variable $p$ in the spacetime momentum integral has been rescaled by a factor of $R$ to be dimensionless. By rotational invariance of Euclidean spacetime, for suitable functions $f(p^2)$ one has\footnote{The same result can be derived rescaling the time component $p_0 \to \alpha \, p_0$ and taking derivatives with respect to $\alpha$.}
\begin{eqaed}\label{eq:connection_trick}
    \int d^dp \, p_0^2 \, f'(p^2) = \int d^dp \, \frac{p^2}{d} \, f'(p^2) = - \, \frac{1}{2} \int d^dp \, f(p^2) \, ,
\end{eqaed}
which for $f'(p^2) \equiv \sum_{n \in \mathbb{Z}^q} \frac{1}{p^2 + n^2} - \int_{\mathbb{R}^q} \frac{d^qn}{p^2 + n^2}$ recovers the (generalization of) ``trace-log'' expression of \cref{eq:eff_pot_circle}. One can then recast this expression as \cref{eq:casimir_torus_heat_kernel} with a Schwinger parametrization. As we will discuss, for deformed geometries it turns out to be more convenient to take this route, since it leads more directly to a simplified expression for the Schwinger integral.

\subsection{Flat deformations are on-shell}\label{sec:casimir_flat_deformations}

Having studied in detail the Casimir energy for square tori, we now include deformations that leave the torus flat. For the purposes of this paper, we do not need to discuss these deformations in full generality; see \cite{ValeixoBento:2025yhz, DallAgata:2025jii} for such an account. In this section we provide some examples showing that flat deformations are on-shell, namely that the Casimir potential is stationary along these directions at the solution of \cite{Luca:2022inb} and its generalizations discussed in \cref{sec:casimir_vacua}.

The simplest example of flat deformation is the case in which one or more radii in the $T^q$ torus are deformed independently. Since the volume (radion) direction is a minimum, we focus on volume-preserving deformations. Let us consider the case where $k < q$ radii are parametrized as 
\begin{eqaed}
    R_1 = \cdots = R_k = R (1+\phi) \, .
\end{eqaed}
Then, in order to get a deformation that is orthogonal to the overall volume, we fix $l$ additional radii to be 
\begin{eqaed}
    R_{k+1} = \cdots = R_{k+l} = R (1+\phi)^{-\frac{k}{l}} \, .
\end{eqaed}
The computation of the Casimir energy is similar to the undeformed case, with the difference that now the zeta function is modified to account for different values of the radii. The lattice sum producing the deformed zeta function takes the form
\begin{eqaed}\label{eq:casimir_def_torus}
    \sum_{n \neq 0} \left( 
(n_1^2 + \cdots + n_k^2)(1+\phi)^2 + \frac{n_{k+1}^2 + \cdots + n_{k+l}^2}{(1+\phi)^{-\frac{2k}{l}}} + (n_{k+l+1}^2 + \cdots + n_q^2) \right)^{-\frac{d+q}{2}} \, .
\end{eqaed}
Dropping uninteresting pre-factors, the derivative of this Casimir energy with respect to $\phi$ at the point $\phi=0$ is proportional to
\begin{eqaed}\label{eq:der_casimir_def_torus}
    \sum_{n \neq 0} \frac{l(n_1^2 + \cdots + n_k^2) - k (n_{k+1}^2 + \cdots + n_{k+l}^2) }{\abs{ \vec{n}}^{d+q+1}} \, ,
\end{eqaed}
which vanishes. To see this, let us we consider the set of terms in this sum associated to $\vec{n}$ that are all permutations of the same unordered collection $C$ of $k+l$ integers. Over each permutation of $C$ the denominator is the same, and will factor out from their sum. In the sum of all such terms, a single element $n_i$ of the collection of integers will show up $k$ times in the first bracket and $l$ times in the second, therefore resulting in a vanishing sum. Notice that we have assumed we can label the single element of the collection $C$. This is clearly true when $C$ is given by $k+l$ different integers, but even in the case where copies of the same number are present the result would always be zero, since the true counting would amount to divide by an irrelevant combinatoric factor. These deformations are hence on-shell.

Another simple example of flat deformation is complex structure deformation for a $T^2$. We can imagine to pick up a $T^2$ factor out of $T^q$ with non-trivial real part of its complex-structure modulus $\operatorname{Re}\tau = \sigma$. The metric on this $T^2$ will be given by
\begin{eqaed}
    ds^2_{T^2} = R^2 (dx + \sigma dy)^2 + R^2 dy^2 \, .
\end{eqaed}
Notice that $\sigma$ defines a deformation orthogonal to the volume of $T^2$, and thus of the complete $T^q$ as well.

The metric can be cast into the standard Euclidean form up to modifying the coordinate periodicities to
\begin{eqaed}
    y &\simeq y + 2\pi m R \, , \\
    x+ \sigma y &\simeq x+ \sigma (y + 2\pi m R) + 2\pi n R \, , \\
    \Longrightarrow x &\simeq x + 2\pi (n + \sigma m) R \, .
\end{eqaed}
The computations that lead to \cref{eq:casimir_torus} is now modified by replacing one of the $n_i$ with $n_i + \sigma n_j$. Without loss of generality we take $i=j-1=1$, which leads to
\begin{eqaed}\label{eq:casimir_CS_torus}
    V_{T^q}(R,\sigma) = - \, \frac{1}{(2\pi R)^{d+q} \, \Omega_{d+q-1}} \, \sum_{n \neq 0} \left(
(n_1+\sigma n_2)^2 + n_2^2 + \cdots + n_q^2) \right)^{-\frac{d+q}{2}} \, .
\end{eqaed}
Performing the derivative of this potential with respect to $\sigma$ at $\sigma=0$ we obtain
\begin{eqaed}\label{eq:casimir_derCS_torus}
    \frac{\partial V}{\partial \sigma}(R,0) =  \, \frac{d+q}{(2\pi R)^{d+q} \, \Omega_{d+q-1}} \, \sum_{n \neq 0} n_2 n_1^2 \left(
n_1^2 + n_2^2 + \cdots + n_q^2) \right)^{-\frac{d+q+2}{2}} \, .
\end{eqaed}
This expression vanishes, since the generic term of the sum is odd under the exchange $n_2 \rightarrow -n_2$. Therefore, the potential is stationary along these directions. In the following section we will prove that indeed all flat deformations are on shell. The method involves computing the leading deviation in the Casimir energy due to a generic deformation that is orthogonal to the volume, whether it is flat or makes the torus curved.

\section{Casimir energy for curved tori}\label{sec:casimir_torus_pert}

We now move on to consider more general deformations of the torus which generate curvature. According to the above considerations, it is enough to consider metric deformations\footnote{In order to be consistent with the preceding discussion, we stick to the conventions in which each internal coordinate has range $y^i \in [0, 2\pi R]$.}
\begin{eqaed}\label{eq:torus_metric}
    ds^2 = \eta_{\mu \nu} \, dx^\mu \otimes dx^\nu + \left( \delta_{ij} + h_{ij}(y) \right) dy^i \otimes dy^j \, ,
\end{eqaed}
where we remove spacetime dependence to retain maximally symmetric solutions. Without loss of generality we restrict to deformations that are orthogonal to the volume, so that $\delta_{ij} h^{ij} = 0$ at the linearized level.

\subsection{Perturbed heat kernel and propagator}\label{sec:heat_kernel_propagator_pert}

In order to find the perturbed heat kernel and propagator, it is useful to employ the schematic notation introduced in \cref{sec:heat_kernel}. Let $L = L_0 + L_1$ be the decomposition of the kinetic operator in terms of the perturbation $L_1$. In the following, $f_i$ and $\lambda_i$ denote eigenfunctions and eigenvalues of the unperturbed operator $L_0$. If the spectrum is non-degenerate, one can easily obtain the perturbed eigenvalues $\lambda_i + \langle f_i \, , \, L_1 \, f_i \rangle$ and linearize the general expressions in \cref{eq:heat_kernel_gen} and \cref{eq:heat_kernel_propagator}, but this is unfortunately not possible in the settings at stake. One would rather diagonalize the operator whose matrix elements are given by
\begin{eqaed}\label{eq:matrix_elements}
    (L_1)_{ij} \equiv \langle f_i \, , \, L_1 \, f_j \rangle = \int dx \, f_i(x)^* \, (L_1 f_j)(x) \, .
\end{eqaed}
Therefore we follow a different strategy, linearizing the defining equations for $K = K_0 + K_1$ and $G = G_0 + G_1$ and solving them directly. The resulting expressions involve the matrix elements $(L_1)_{ij}$ without the need for diagonalization.

\noindent With the notation of \cref{eq:torus_metric}, the unperturbed operator on $\mathbb{R}^{d-1,1} \times T^q_R$ is the standard d'Alembertian $L_0 = \Box = \partial_\mu \partial^\mu + \partial_i \partial^i$. The perturbation is the linearized Laplace-Beltrami operator, which for traceless deformations $h_{ij}$ simplifies to
\begin{eqaed}\label{eq:perturbed_laplacian}
    L_1 = - \, \partial_i (h^{ij} \, \partial_j) \, .
\end{eqaed}
The complete set $\{f_{p,n}(x,y)\} = \{e^{i p \cdot x + i \frac{n}{R}\cdot y} \}$ is specified by the quantum numbers $p$ and $n$, and is (distributionally) normalized such that
\begin{eqaed}\label{eq:orthogonal_normalization}
    \langle f_{p,n} \, , \, f_{q,m}\rangle = \int_{\mathbb{R}^{d-1,1} \times T^q} \!\!\!\!\!\!\!\!\!\!\!\!\!\!\!\!\!\! d^dx \, d^qy \, e^{i(p-q) \cdot x + i\frac{n-m}{R} \cdot y} = (2\pi)^{d+q}R^q \, \delta^{(d)}(p-q) \, \delta_{n,m} \, .
\end{eqaed}
Thus, the matrix elements of the perturbed operator are given by
\begin{eqaed}\label{eq:L1_matrix_elements}
    (L_1)_{(p,n)(q,m)} = (2\pi)^d \, \delta^{(d)}(p-q) \, \frac{1}{R^2} \, n_i \, m_j \, \widetilde{h}^{ij}_{n-m} \, ,
\end{eqaed}
where $\widetilde{h}_n \equiv \int_{T^q} d^qy \, h(y) \, e^{i \frac{n}{R} \cdot y}$ are the Fourier modes of the metric perturbation.

\noindent With these ingredients at hand, we can derive the general form of the corrections $K_1$ and $G_1$ to the heat kernel and propagator. The linearized heat-kernel equation reads
\begin{eqaed}\label{eq:linearized_heat_kernel}
    (\partial_s + L_0) \, K_1 = - \, L_1 K_0 \, .
\end{eqaed}
Projecting onto the spectral basis $\{f_i(x)\}$, writing $K_1 = \sum_i f_i(x)^* \, k_i(y \, | \, s)$, one obtains
\begin{eqaed}\label{eq:linearized_heat_kernel_proj}
    (\partial_s + \lambda_i) \, k_i = - \sum_j (L_1)_{ij} \, f_j(y) \, e^{-\lambda_j s} \, . 
\end{eqaed}
Solving this equation by an integrating factor, in the sum over $j$ one must distinguish whether $\lambda_i = \lambda_j$, since the two primitives differ. Furthermore, one must impose the initial condition $K_1 \overset{s \to 0^+}{\to} 0$. The final expression is
\begin{eqaed}\label{eq:linearized_heat_kernel_sol}
K_1 = \sum_{i, j} (L_1)_{ij} \, f_i(x)^* \, f_j(y) \left( \frac{e^{-\lambda_i s} - e^{-\lambda_j s}}{\lambda_i - \lambda_j}\bigg|_{\lambda_i \neq \lambda_j} - \, s \, e^{-\lambda_i s}\bigg|_{\lambda_i = \lambda_j} \right) \, , 
\end{eqaed}
where the first term restricts the sum to $\lambda_i \neq \lambda_j$ and the second to $\lambda_i = \lambda_j$. This is the main reason why it is more convenient to use the propagator: writing $G_1 = - \int_0^\infty ds \, K_1$, the two contributions add up to the simple result
\begin{eqaed}\label{eq:linearized_prop_from_heat_kernel}
    G_1 = \sum_{ij} (L_1)_{ij} \, \frac{f_i(x)^* \, f_j(y)}{\lambda_i \lambda_j} \, .
\end{eqaed}
One finds the same result directly solving the linearized equation for the propagator\footnote{When applying this method for deformations that are not orthogonal to the volume, one must include the variation of the $\delta$ term which includes a metric determinant.},
\begin{eqaed}\label{eq:linearized_propagator_eq}
    L_0 \, G_1 = - \, L_1 \, G_0 \, .
\end{eqaed}
As we have anticipated, we found it simpler to employ \cref{eq:linearized_prop_from_heat_kernel} rather than \cref{eq:linearized_heat_kernel_sol}, introducing a Schwinger parameter in a different way.

\noindent For the setting at hand, using \cref{eq:L1_matrix_elements} one arrives at
\begin{eqaed}\label{eq:linearized_propagator_torus}
    G_1((x,y), (x',y')) = \int \frac{d^dp}{(2\pi)^d} \sum_{n, m \in r\mathbb{Z}^q} \frac{1}{(2\pi R)^{2q}} \, \frac{n_i \, m_j}{R^2} \, \frac{\widetilde{h}^{ij}_{n-m} \, e^{-i p\cdot (x - x') - i \frac{n \cdot y - m \cdot y'}{R} }}{\left(p^2 + \frac{n^2}{R^2}\right)\left(p^2 + \frac{m^2}{R^2}\right)} \, .
\end{eqaed}
The sum can go over the integers ($r=1$) or over the half integers ($r=1/2$), depending on the boundary conditions on the torus circles. We can now use this formula to compute the perturbed Casimir energy density to first order. The unregulated contribution reads
\begin{equation}\label{eq:perturbed.T00}
	\delta V = - \int \frac{d^dp}{(2\pi)^d} \frac{p^2}{d} \sum_{n, m \in r\mathbb{Z}^q} \frac{1}{(2\pi R)^{2q}} \, \frac{n_i \, m_j}{R^2} \, \frac{\widetilde{h}^{ij}_{n-m} \, e^{- i \frac{(n-m) \cdot y}{R} }}{\left(p^2 + \frac{n^2}{R^2}\right)\left(p^2 + \frac{m^2}{R^2}\right)} \, ,
\end{equation}
whose ultraviolet divergences are due to local curvature invariants canceled by the corresponding contribution on $\mathbb{R}^{d-1,1} \times \mathbb{R}^q$ equipped with the same metric.

Although we will not need it in the following, we also provide the second-order expressions for the heat-kernel trace, from which one can extract the second-order contribution (and thus the masses) to the Casimir energy for any deformations, flat or curved. Letting $L = L_0 + L_1 + L_2$ denote the deformed kinetic operator up to second order, the trace $\text{tr} K_2$ of the second-order correction to the heat kernel is given by
\begin{eqaed}\label{eq:trK2_1}
    \text{tr} K_2 & =  \sum_{i,j} (L_1)_{ij}(L_1)_{ji} \left[\left( \frac{e^{-\lambda_i s}{ - e^{-\lambda_j s}}}{(\lambda_i - \lambda_j)^2} - \frac{s \, e^{-\lambda_i s}}{\lambda_i - \lambda_j} \right)_{\lambda_i \neq \lambda_j} \hspace{-12pt} + \frac{s^2}{2} \, e^{-\lambda_i s}\bigg|_{\lambda_i = \lambda_j} \right] \\
    & - s \sum_i (L_2)_{ii} \, e^{-\lambda_i s} \, .
\end{eqaed}
The first term in the round brackets contributes zero to the result for symmetry reasons, whereas the second can be written in a symmetric way in $i$ and $j$. This simplifies the expression to
\begin{eqaed}\label{eq:trK2_2}
    \text{tr} K_2 = -s \sum_i (L_2)_{ii} e^{-\lambda_i s} - \frac{s}{2} \sum_{i,j} (L_1)_{ij}(L_1)_{ji} \left(\frac{e^{-\lambda_i s} - e^{-\lambda_j s}}{\lambda_i - \lambda_j} \bigg|_{\lambda_i \neq \lambda_j} \hspace{-25pt}- s \, e^{-\lambda_i s}\bigg|_{\lambda_i = \lambda_j} \right) .
\end{eqaed}
This leads to the following expressions for the potential and the (trace of the) propagator,
\begin{eqaed}\label{eq:trG_V_2nd-order}
    - \int_0^\infty ds \, \text{tr} K_2 & = \sum_i \frac{(L_2)_{ii}}{\lambda_i^2} - \frac{1}{2} \sum_{i,j} (L_1^2)_{ij} \, \frac{\lambda_i + \lambda_j}{\lambda_i^2 \lambda_j^2} \, , \\
    - \int_0^\infty \frac{ds}{s} \, \text{tr} K_2 & = \sum_i \frac{(L_2)_{ii}}{\lambda_i} - \frac{1}{2} \sum_{i,j} \frac{(L_1^2)_{ij}}{\lambda_i \lambda_j} \, .
\end{eqaed}

\section{Absence of tadpoles}\label{sec:tadpoles}

The result in (\ref{eq:perturbed.T00}) encodes the first functional derivative of the full potential, and readily implies that flat deformations, with constant $h$, are on-shell. To wit, in this case $\widetilde{h}^{ij}_{n-m} = (2\pi R)^{q} \, \delta_{n,m} \, h^{ij}$, resulting in the combination $n_i \, n_j \, h^{ij}$ in the remaining sum over $n$. Since all other factors only contain $n^2$, by rotational invariance the result is proportional to the trace of $h$, which vanishes. This shows in full generality that flat deformations are indeed on shell, as discussed in the explicit examples above.

In general, this will not necessarily occur for deformations that depend on the internal coordinates. However, it turns out that it is enough that contributions from flat deformations vanish for the geometry to be on-shell, since linear terms for the modes encoding curvature corrections vanish when integrated over the internal space. Hence, they do not contribute to the variation of the action even before dimensional reduction. Indeed, corrections to the potential take the schematic form $\delta V=\sum_{\vec{m}} a_{\vec{m}} h_{\vec{m}}e^{i\frac{\vec{m}\cdot \vec{y}}{R}}$, for some coefficients $a_{\vec{m}}$, which gives rise to a term in the action of the form
\begin{equation}\label{eq:tadpoles}
	\int d^{d+q} x \sum_{\vec{m}} a_{\vec{m}} h_{\vec{m}} e^{i\frac{\vec{m}\cdot \vec{y}}{R}} = \sum_{\vec{m}} a_{\vec{m}} \int d^{d+q} x \, h_{\vec{m}} e^{i\frac{\vec{m}\cdot \vec{y}}{R}} \, .
\end{equation}
Performing the integral in the internal coordinates, one finds that this term is proportional to the coefficient $a_{\vec{0}}$ accompanying the flat deformations, which vanishes. This is consistent with the expectation that integrating out matter fields commutes with taking the gravitational field equations, the latter being the procedure of \cite{Luca:2022inb} to arrive at the solutions presented in \cref{sec:casimir_vacua}. Indeed, for matter fluctuations, the expectation value of the energy-momentum tensor used in the semiclassical field equations is the metric variation of the matter effective action we computed. This reasoning does not quite cover graviton fluctuations, for which there is no standard energy-momentum tensor; however, perturbatively it is possible to separate background and fluctuations, and gauge-fix small diffeomorphisms.

At any rate, for the above argument to work, the corrections need to be well-defined. That is, for every choice of deformation, the series appearing (\ref{eq:tadpoles}) ought to be (uniformly) convergent. This will be the case provided that the sequence $\{a_{\vec{m}}\}$ is square-summable. A proof of this statement is provided in the next section.

\subsection{Proof of existence of the functional derivative}\label{sec:well.def.casimir}

We begin by noting that it is enough to check convergence for a linearly independent set of traceless perturbations, since (\ref{eq:perturbed.T00}) is a linear functional of $h^{ij}$. For the moment, let us focus on traceless diagonal perturbations. Any such metric deformation can be decomposed into a sum of perturbations of the form $h^{ii} = -h^{jj}$, where $i \neq j$. We specialize to the case $i = 1$, $j = 2$, as all others are analogous. Then, the perturbation can be expanded in a Fourier series as
\begin{equation}
h^{11} = -h^{22} = \sum_{\vec{m}} h_{\vec{m}} \, e^{i\vec{m} \cdot \vec{y}/R} \, ,
\end{equation}
where $h_{\vec{m}}$ is a sequence in $\ell^2(\mathbb{Z}^q, \mathbb{C})$. The corrections for a single field take the form $\delta V=\sum_{\vec{m}}\delta V_{\vec{m}}$, where
\begin{equation}\label{eq:individual.corrections}
    \delta V_{\vec{m}}  = - \, \frac{h_{\vec{m}} \, e^{i\vec{m}\cdot\vec{y}/R}}{(2\pi R)^{d+q}} \, \int d^dp \, \frac{p^2}{d} \, \sum_{\vec{n}\in r\mathbb{Z}^{q}} \frac{n_1(n_1+m_1) - n_2(n_2+m_2)}{\left(p^2 + (\vec{n}+\vec{m})^2\right)\left(p^2 + \vec{n}^2\right)} \, .
\end{equation}
It is useful to rewrite this expression using a Schwinger-like parameter via the identity
\begin{equation}
    \frac{1}{ab} = \frac{1}{b-a}\int_0^\infty ds \left( e^{-as} - e^{-bs}\right)
\end{equation}
applied to the denominator inside the sum. The (dimensionless) momentum integrals then give
\begin{equation}
    \int d^dp \, \frac{p^2}{d} \, e^{-p^2 s} = \frac{\Omega_{d-1}}{4} \, \Gamma\left(\frac{d}{2}\right) \, s^{- \frac{d}{2} - 1} = \frac{\pi^{\frac{d}{2}}}{2 s^{\frac{d}{2}+1}} \, ,
\end{equation}
and thus \cref{eq:individual.corrections} simplifies to
\begin{equation}\label{eq:diagonal}
    a_{\vec{m}}=- \, \frac{\pi^{\frac{d}{2}} \,h_{\vec{m}} \, e^{i\vec{m}\cdot\vec{y}/R}}{2(2\pi R)^{d+q}} \int_0^\infty \frac{ds}{s^{\frac{d}{2}+1}} \sum_{\vec{n} \in r\mathbb{Z}^q} \frac{n_1(n_1+m_1)-n_2(n_2+m_2)}{\vec{m}\cdot(2\vec{n}+\vec{m})} \left( e^{-\vec{n}^2 s} - e^{-(\vec{n}+\vec{m})^2 s} \right) .
\end{equation}
The integral over the Schwinger parameter is divergent. Similar to the divergence encountered in the first term of the expansion, this behavior arises from the non-compactness of the background and the physical Casimir energy is given by subtracting the corresponding uncompactified contribution. Actually, in the settings at stake, due to the equal number of bosonic and fermionic degrees of freedom the full vacuum energy does not feature this \ac{UV}-sensitive issue to begin with. Hence, this subtraction can be consistently applied to each term without altering the physical result that couples to the gravitational field. We will now show that each individual term, when combined with the flat space contribution, is finite; furthermore, we will prove that for any square-summable metric perturbation the sum of these contributions is also finite. 

Let us write the first-order corrections as
\begin{equation}
    \delta V_{\vec{m}} = a_{\vec{m}}e^{-i\vec{m}\cdot\vec{y}/R} \, .
\end{equation}
The coefficients $a_{\vec{m}}$ can be read off from (\ref{eq:individual.corrections}) and take the form
\begin{equation}\label{eq:diagonal_2}
    a_{\vec{m}}=- \, \frac{\pi^{\frac{d}{2}} \,h_{\vec{m}} \, e^{i\vec{m}\cdot\vec{y}/R}}{2(2\pi R)^{d+q}} \int_0^\infty \frac{ds}{s^{\frac{d}{2}+1}}  \left ( S_{\vec{m}}-J_{\vec{m}}\right) ,
\end{equation}
where we have defined
\begin{equation}
    S_{\vec{m}} = \sum_{\vec{n} \in r\mathbb{Z}^q} \frac{n_1(n_1+m_1)-n_2(n_2+m_2)}{\vec{m}\cdot(2\vec{n}+\vec{m})} \left( e^{-\vec{n}^2 s} - e^{-(\vec{n}+\vec{m})^2 s} \right) ,
\end{equation}
which encodes the torus contribution, and its uncompactified counterpart
\begin{equation}
    J_{\vec{m}} = \int d^q n\,\frac{n_1(n_1+m_1)-n_2(n_2+m_2)}{\vec{m}\cdot(2\vec{n}+\vec{m})} \left( e^{-\vec{n}^2 s} - e^{-(\vec{n}+\vec{m})^2 s} \right).
\end{equation}
In order to analyze convergence of the Schwinger integral, we do the following. First, we recast $S_{\vec{m}}$ in a more symmetric form by performing the change of variables $\vec{n}\rightarrow \vec{n}-\vec{m}/2$, obtaining
\begin{equation}
    S_{\vec{m}} = \sum_{\vec{n} \in r\mathbb{Z}^q} \frac{n_1^2-n_2^2-(m_1^2-m_2^2)/4}{2 \vec{m}\cdot\vec{n}} \left( e^{-(\vec{n}-\vec{m}/2)^2 s} - e^{-(\vec{n}+\vec{m}/2)^2 s} \right) .
\end{equation}
Next, we introduce a real parameter $a\in[0,1]$ and define
\begin{equation}
    S_{(\vec{m},a)} = \sum_{\vec{n} \in r\mathbb{Z}^q} \frac{n_1^2-n_2^2-(m_1^2-m_2^2)/4}{2 \vec{m}\cdot\vec{n}} \left( e^{-(\vec{n}-a\vec{m}/2)^2 s} - e^{-(\vec{n}+a\vec{m}/2)^2 s} \right) .
\end{equation}
This expression satisfies $S_{(\vec{m},1)}=S_{\vec{m}}$ and $S_{(\vec{m},0)}=0$ and, additionally, it obeys the differential equation
\begin{equation}\label{eq:fulldiffeq}
    \partial_a S_{(\vec{m},a)}+\frac{\vec{m}^2 s}{2}a\,S_{(\vec{m},a)}= s\sum_{\vec{n} \in r\mathbb{Z}^q}(n_1^2-n_2^2-(m_1^2-m_2^2)/4)e^{-(\vec{n}+a\vec{m}/2)^2 s} \, .
\end{equation}
The original $S_{\vec{m}}$ is thus the solution with initial condition $S_{(\vec{m},0)}=0$ evaluated at $a=1$. To proceed, we use the identity $n^2 = (n+am/2)^2-am(n+am/2)+a^2m^2/4$ and rewrite the right-hand side in the following way,
\begin{equation}
\begin{split}
    s\sum_{\vec{n}} (n_1+am_1/2)^2 e^{-(\vec{n}+a\vec{m}/2)^2s} - sam_1\sum_{\vec{n}} (n_1+am_1/2) e^{-(\vec{n}+a\vec{m}/2)^2s} &\\+ s(a^2-1)\frac{m_1^2}{4} \sum_{\vec{n}} e^{-(\vec{n}+a\vec{m}/2)^2s} 
  - s\sum_{\vec{n}} (n_2+am_2/2)^2 e^{-(\vec{n}+a\vec{m}/2)^2s} + &\\ sam_2\sum_{\vec{n}} (n_2+am_2/2) e^{-(\vec{n}+a\vec{m}/2)^2s} - s(a^2-1)\frac{m_2^2}{4} \sum_{\vec{n}} e^{-(\vec{n}+a\vec{m}/2)^2s} \, . &
\end{split}
\end{equation}
We now split $S_{(\vec{m},a)}$ into six parts, so that each one of them solves \cref{eq:fulldiffeq} for each of the inhomogeneous terms above. Specifically, we write
\begin{equation}
    S_{(\vec{m},a)} = \sum_{i=1}^{6} S_{i(\vec{m},a)} 
\end{equation}
with initial conditions $S_{i(\vec{m},0)} = 0$, which ensures $S_{(\vec{m},0)} = 0$. This decomposition is uniquely determined since the solution to the initial value problem in \cref{eq:fulldiffeq} is unique. For the flat-space contribution, we similarly define
\begin{equation}
    J_{(\vec{m},a)} = \int d^q n\,\frac{n_1^2-n_2^2-(m_1^2-m_2^2)/4}{2 \vec{m}\cdot\vec{n}} \left( e^{-(\vec{n}-a\vec{m}/2)^2 s} - e^{-(\vec{n}+a\vec{m}/2)^2 s} \right).
\end{equation}
This obeys a differential equation analogous to (\ref{eq:fulldiffeq}), where summations are replaced by integrals,
\begin{align}\label{eq:fulldiffeq_2}
    \partial_a J_{(\vec{m},a)}+\frac{\vec{m}^2 s}{2}a\,J_{(\vec{m},a)} 
    &= s\int d^qn\left(n_1^2 - n_2^2 - \frac{m_1^2 - m_2^2}{4}\right)
    e^{-(\vec{n}+a\vec{m}/2)^2 s} \nonumber \\
    &= s(a^2-1)\frac{m_2^2 - m_1^2}{4} \frac{1}{2} \left(\frac{\pi}{s}\right)^{q/2}.
\end{align}
Analogously, we decompose it in the same manner as $S_{(\vec{m},a)}$,
\begin{equation}
    J_{(\vec{m},a)} = \sum_{i=1}^{6} J_{i(\vec{m},a)} \, .
\end{equation}
For convenience, let $S_{i,\vec{m}}:=S_{i(\vec{m},1)}$ and $J_{i,\vec{m}}:=J_{i(\vec{m},1)}$. Then, $a_{\vec{m}}$ can be expressed as
\begin{equation}
    a_{\vec{m}} = -\frac{\pi^{d/2}h_{\vec{m}} }{2(2\pi R)^{d+q}} 
    \int_0^\infty \frac{ds}{s^{d/2+1}} \sum_{i=1}^6 \left(S_{i,\vec{m}} - J_{i,\vec{m}}\right).
\end{equation}
We analyze now each of the six terms separately.

\begin{figure}[htbp]
    \centering
    \includegraphics[width=0.7\textwidth]{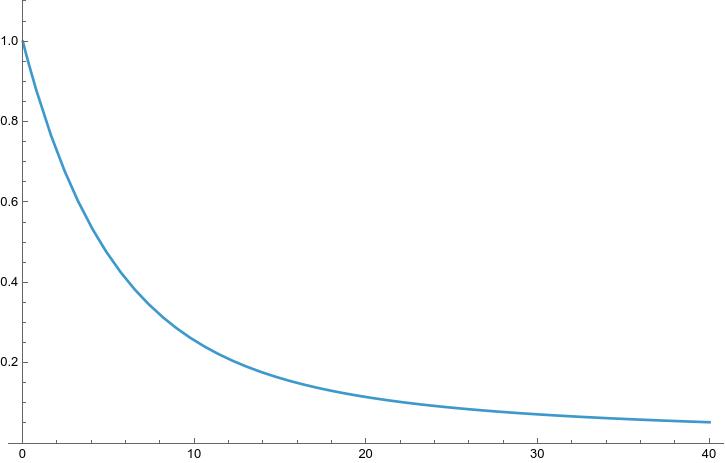}  
    \caption{Plot of $F(x)$ defined in \cref{eq:F_fun}.}
    \label{fig:F}
\end{figure}

We start with $S_{1,\vec{m}}-J_{1,\vec{m}}$ and $S_{4,\vec{m}}-J_{4,\vec{m}}$ . We focus on the case $i=1$, but the same argument applies to $i=4$, since their inhomogeneous terms only differ by a sign. We have the differential equation
\begin{equation}
    \left (\partial_a+\frac{\vec{m}^2 s}{2}a\right )\,(S_{1,(\vec{m},a)}-J_{1,(\vec{m},a)})= s\sum_{\vec{n}} (n_1+am_1/2)^2 e^{-(\vec{n}+a\vec{m}/2)^2s} - \frac{1}{2}\left (\frac{\pi}{s}\right)^{q/2}.
\end{equation}
By performing a Poisson resummation, the right-hand side can be rewritten according to\footnote{Inside the cosine in the second line of \cref{eq:bound.S14} there can be an extra phase associated with the fact that the sum goes over the half integers for fermionic contributions. This phase is irrelevant, since at any rate we bound the cosine by unity. This remark also applies to other bounds we will derive later.} 
\begin{equation}\label{eq:bound.S14}
\begin{split}
    & s\sum_{\vec{n}} (n_1+am_1/2)^2 e^{-(\vec{n}+a\vec{m}/2)^2s} - \frac{1}{2}\left(\frac{\pi}{s}\right)^{q/2} \\
    & \quad = 2^{q}\left(\frac{\pi}{s}\right)^{\frac{q}{2}}\sum_{k_i=1}^{\infty} \left( \frac{k_1^2\pi^2}{s}-\frac{1}{2} \right)\prod_{i=1}^{q}\cos (2\pi am_ik_i)e^{-\frac{ k_i^2\pi^2}{s}}.
\end{split}
\end{equation}
With this in mind, we take the absolute value in both sides of the differential equation and bound
\begin{equation}
\begin{split}
    \left| \left(\partial_a + \frac{\vec{m}^2 s}{2}a\right)(S_{1,(\vec{m},a)} - J_{1,(\vec{m},a)}) \right|
    &\leq 2^{q}\left(\frac{\pi}{s}\right)^{\frac{q}{2}}\sum_{k_i=1}^{\infty} \left(\frac{k_1^2\pi^2}{s} + 1\right)e^{-\frac{\sum_i k_i^2\pi^2}{s}} \\
    &:= f_1(s).
\end{split}
\end{equation}
This function $f_1(s)$ has the property that
\begin{equation}\label{eq:f.property}
    \int_0^\infty \frac{ds}{s^{d/2+1}}f_1(s) <\infty.
\end{equation}
for $d>0$. This is because, for small $s$, $f_1$ behaves like $\mathcal{O}(e^{-\pi^2/s})$, while for large $s$ one can check (using, for example, the integral test) that $f_1(s)$ cannot grow faster than a constant. Finally, we can safely integrate both sides in $a$ while preserving the bound and find
\begin{equation}
\left |S_{1,\vec{m}}-J_{1,\vec{m}}\right | \leq f_1(s) \,F(\vec{m}^2s)  \leq f_1(s) , 
\end{equation}
where we have defined
\begin{equation}\label{eq:F_fun}
    F(x) = \frac{2e^{-x/4}}{\sqrt{x}} \int_0^{\sqrt{x}/2} e^{y^2}dy \,.
\end{equation}
This function is plotted in \cref{fig:F}. Of course, this bound is also valid for $\left |S_{4,\vec{m}}-J_{4,\vec{m}}\right |$.

In the case of $S_{2,\vec{m}}-J_{2,\vec{m}}$ (and also  $S_{5,\vec{m}}-J_{5,\vec{m}}$), we first note that
\begin{equation}
\begin{split}
     \sum_{\vec{n}}(n+am/2) e^{-(n+am/2)^2s} = \frac{2}{s}\sqrt{\frac{\pi}{s}}\sum_{k=1}^{\infty} k \sin (am \pi k) e^{-\frac{k^2\pi}{s}}.
\end{split}
\end{equation}
We take absolute values in the differential equation and bound
\begin{equation}
\begin{split}
    \left |\left(\partial_a + \frac{\vec{m}^2 s}{2}a\right)(S_{2,(\vec{m},a)} - J_{2,(\vec{m},a)}) \right | 
    &=\left | asm_1\sum_{\vec{n}}(n_1+am_1/2) e^{-(\vec{n}+a\vec{m}/2)^2s} \right | \\
    &\leq  \frac{1}{2}a s |m_1| \, \theta_3^{q-1}\frac{4}{s}\sqrt{\frac{\pi}{s}} \sum_{k=1}^{\infty} k e^{-\frac{k^2 \pi^2}{s}} \\
    &:= \frac{1}{2} as|m_1| f_2(s),
\end{split}
\end{equation}
where $f_2$ is a function that also satisfies\footnote{Since the general term in the sum $\sum_{k=1}^{\infty} \frac{k}{s} e^{-\frac{k^2 \pi^2}{s}}$ is not decreasing for all $k$ and the region where it is increasing gets larger and larger with $s$, one may worry that the integral test cannot be used to assert that $f_2$ is uniformly bounded by a constant. A way out is to consider a more generous bound $f_2(s) < 4\theta_3^{q-1}\sqrt{\frac{\pi}{s}}\sum^{\infty}_{k=1}\left (\frac{k}{s} + 1\right )e^{-\frac{k^2 \pi^2}{s}}$. The general term in the sum is now decreasing for all $k\geq1$, and $f_2$ grows at most like a constant for large $s$.} \cref{eq:f.property}. Then, we find that
\begin{equation}
    |S_{2,\vec{m}}-J_{2,\vec{m}}|\, \leq \,  f_2(s)\frac{|m_1|}{\vec{m}^2} G(\vec{m}^2s) \leq f_2(s) \, .
\end{equation}
Here $G(x):=1-e^{-x/4}$, and we used $G(x)\leq1$. Note that the bound also applies to $|S_{5,\vec{m}}-J_{5,\vec{m}}|$.

\begin{figure}[htbp]
    \centering
    \includegraphics[width=0.7\textwidth]{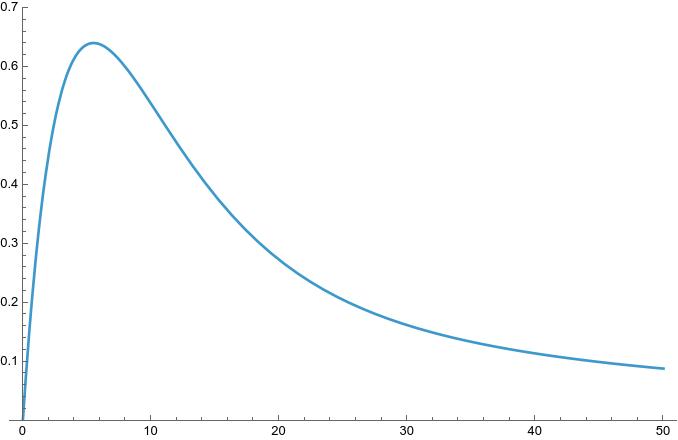}  
    \caption{Plot of $H(x)$ defined in \cref{eq:H_fun}.}
    \label{fig:H}
\end{figure}

Lastly, we do the same for $S_{3,\vec{m}}-J_{3,\vec{m}}$ and $S_{6,\vec{m}}-J_{6,\vec{m}}$. The differential equation is 
\begin{equation}
\begin{split}
    \left(\partial_a + \frac{\vec{m}^2 s}{2}a\right)(S_{3,(\vec{m},a)} - J_{3,(\vec{m},a)}) 
    &=(a^2-1)\frac{m_1^2s}{4}\left (\sum_{\vec{n}} e^{-(\vec{n}+a\vec{m}/2)^2s} - \left(\frac{\pi}{s}\right)^{\frac{q}{2}} \right ) \\
    &= (a^2-1)\frac{m_1^2s}{4} 2^{q}\left(\frac{\pi}{s}\right)^{\frac{q}{2}}\prod_{i=1}^q\sum_{k_i=1}^\infty \cos(am_i k_i\pi)\,e^{-\frac{k_i^2\pi^2}{s}}.
\end{split}
\end{equation}
Once again, we take absolute values and bound
\begin{equation}
\begin{split}
    \left |\left(\partial_a + \frac{\vec{m}^2 s}{2}a\right)(S_{3,(\vec{m},a)} - J_{3,(\vec{m},a)})\right | 
    &=(1-a^2)\frac{m_1^2s}{4} 2^{q}\left(\frac{\pi}{s}\right)^{\frac{q}{2}}\left|\prod_{i=1}^q\sum_{k_i=1}^\infty \cos(am_i k_i\pi)\,e^{-\frac{k_i^2\pi^2}{s}}\right|\\
    &\leq (1-a^2)\frac{m_1^2s}{2} 2^{q+1}\left(\frac{\pi}{s}\right)^{\frac{q}{2}}\sum_{k_i=1}^\infty \, e^{-\frac{\sum_i k_i^2\pi^2}{s}} \\&:= (1-a^2)\frac{m_1^2s}{2} f_3(s).
\end{split}
\end{equation}
The function $f_3$ also satisfies \cref{eq:f.property}. After integrating, one obtains
\begin{equation}\label{eq:bound3}
    \left |S_{3,(\vec{m},a)} - J_{3,(\vec{m},a)}\right| \leq  f_3(s) \frac{m_1^2}{\vec{m}^2}H(\vec{m}^2s)\leq f_3(s),
\end{equation}
where $H(x)$ is given by
\begin{equation}\label{eq:H_fun}
     H(x):=\frac{1}{\sqrt{x}}(2+x) e^{-x/4}\int_0^{\sqrt{x}/2} e^{y^2}dy-1 \, .
\end{equation}
As depicted in \cref{fig:H}, $H(x)\leq 1$. Analogously to the preceding cases, the bound also applies to $\left |S_{6,(\vec{m},a)} - J_{6,(\vec{m},a)}\right|$.

We are now in a position to bound the whole coefficient $a_{\vec{m}}$. We have that
\begin{equation}
\begin{split}
    \left| a_{\vec{m}}\right| 
    &\leq \frac{\pi^{d/2} h_{\vec{m}} }{2(2\pi R)^{d+q}} 
       \int_0^\infty \frac{ds}{s^{d/2+1}} \sum_{i=1}^6 \left|S_{i,\vec{m}} - J_{i,\vec{m}}\right| \\
    &\leq \frac{\pi^{d/2} h_{\vec{m}}}{(2\pi R)^{d+q}}
       \int_0^\infty \frac{ds}{s^{d/2+1}} \sum_{i=1}^3 \left|f_i(s)\right|= C h_{\vec{m}} \, .
\end{split}
\end{equation}
Importantly, in the above expression $C$ is a finite positive $\vec{m}$-independent constant. Hence, since $h_{\vec{m}}$ is square-summable, so is $a_{\vec{m}}$. This concludes the proof that the correction to the potential $\delta V$ is finite for diagonal metric perturbations.

We also have to show that off-diagonal perturbations do not yield a divergent series. These are of the form $h^{ij}=h^{ji}$, where $i\neq j$. Once again, it is enough to check this for $i=1$, $j=2$. As before, we expand in Fourier modes
\begin{equation}
    h^{12}=h^{21}= \sum_{\vec{m}}h_{\vec{m}}\,e^{i\vec{m}\cdot \vec{y}} \, .
\end{equation}
The correction takes the form $\delta V=\sum_{\vec{m}}\delta V_{\vec{m}}$, where
\begin{equation}\label{eq:offdiagonal}
    - \frac{\pi^{d/2} h_{\vec{m}} e^{i\vec{m}\cdot\vec{y}/R}}{2(2\pi R)^{d+q}} 
    \int_0^\infty \frac{ds}{s^{d/2+1}} 
    \left (T_{\vec{m}}-K_{\vec{m}}\right) .
\end{equation}
Here we have defined
\begin{equation}
    T_{\vec{m}}= \sum_{\vec{n} \in \mathbb{Z}^q} \frac{n_1(n_2+m_2)+n_2(n_1+m_1)}{\vec{m}\cdot(2\vec{n}+\vec{m})} \left( e^{-\vec{n}^2 s} - e^{-(\vec{n}+\vec{m})^2 s} \right) ,
\end{equation}
and $K_{\vec{m}}$ is the corresponding flat-space contribution. Again, we perform the change of variables $\vec{n}\rightarrow \vec{n}-\vec{m}/2$ and introduce a parameter $a$, so that
\begin{equation}
     T_{(\vec{m},a)}= \sum_{\vec{n} \in \mathbb{Z}^q} \frac{n_1n_2-m_1m_2/4}{\vec{m}\cdot \vec{n}} \left( e^{-(\vec{n}-a\vec{m}/2)^2 s} - e^{-(\vec{n}+a\vec{m}/2)^2 s} \right) .
\end{equation}
The relevant differential equation now reads
\begin{equation}
    \partial_a T_{(\vec{m},a)} + \frac{\vec{m}^2s}{2}a\,T_{(\vec{m},a)} = s\sum_{\vec{n}} \left (n_1n_2-\frac{m_1m_2}{4}\right )e^{-(\vec{n}+a\vec{m}/2)^2s} \, .
\end{equation}
The right hand side can also be written as 
\begin{equation}
\begin{split}
    &s\sum_{\vec{n}} \left(n_1 + a \frac{m_1}{2}\right)\left(n_2 + a \frac{m_2}{2}\right) e^{-(\vec{n} + a\frac{\vec{m}}{2})^2 s} 
    - \frac{a sm_1 }{2} \sum_{\vec{n}} \left(n_2 + a \frac{m_2}{2}\right) e^{-(\vec{n} + a\frac{\vec{m}}{2})^2 s} \\
    &- \frac{a sm_2}{2} \sum_{\vec{n}} \left(n_1 + a \frac{m_1}{2}\right) e^{-(\vec{n} + a\frac{\vec{m}}{2})^2 s} 
    + s(a^2 - 1)\frac{m_1 m_2}{4} \sum_{\vec{n}} e^{-(\vec{n} + a\frac{\vec{m}}{2})^2 s} \, .
\end{split}
\end{equation}
As for diagonal perturbations, we split $T_{(\vec{m},a)}$ into four parts,
\begin{equation}
    T_{(\vec{m},a)}=\sum_{i=1}^4 T_{i(\vec{m},a)} \, ,
\end{equation}
and we do the same for the flat-space contribution, which we now dub $K_{(\vec{m},a)}$. The terms $T_{i,m} - K_{i,m}$, with $i=1,2,3$, have a vanishing flat-space contribution, and they can be bounded in a similar way as $S_{2,\vec{m}}-J_{2,\vec{m}}$. For $T_{4,\vec{m}} - K_{4,\vec{m}}$ the bound is the same as in \cref{eq:bound3}, up to replacing $m_1^2$ with $|m_1m_2|$ in the intermediate step.

All in all, we have argued that no matter what perturbation one considers, the functional derivative of the potential is well-defined. This is true for all dimensions $d$ and $q$. As a consequence, the argument discussed in the beginning of this section is valid, and Casimir vacua of this type are on-shell for all moduli.

\section{Presence of instabilities}\label{sec:tachyons}

So far, we have found that this family of Casimir vacua is on-shell, not just for volume deformations, but also for all other moduli of the torus. It remains to elucidate whether these vacuum solutions are stable. Due to the absence of supersymmetry, at least a universal decay channel of these vacua through brane nucleation is to be expected \cite{Horowitz:2007pr, Brown:2010bc, Brown:2010mf, Brown:2010mg, Ooguri:2016pdq, Antonelli:2019nar, Dibitetto:2020csn, Bomans:2021ara, Dibitetto:2022rzy}. We will explore this in \cref{sec:general_decay}. The more pressing issue is whether the extremum is a \emph{bona fide} classically stable vacuum of the effective potential, namely whether there are directions in field space that give rise to tachyons violating the \ac{BF} bound. In the following section we address this question, finding a tachyonic metric deformation. To this end, we can safely focus on deformations that are classically massless, since Kaluza-Klein modes are parametrically heavier than what the Casimir contribution can destabilize via mixings or mass corrections. Indeed, the leading contribution to the masses is the same as in the supersymmetric case with periodic conditions for the spinors, which ensures that the corrected masses are safely above the stability threshold for these modes. The presence of a supersymmetric counterpart to the solution, although in this case a Minkowski one, is crucial for this argument; indeed, as discussed in \cite{Basile:2018irz}, there exist parametrically controlled Freund-Rubin solutions of non-supersymmetric strings where some classical Kaluza-Klein masses already generate perturbative instabilities.

\subsection{Perturbative instabilities}\label{sec:stability}

We begin by writing down the eleven-dimensional effective potential including deformations that are orthogonal to the volume. Since the fluxes thread the entirety of the internal manifold, their contribution to the potential is only sensitive to volume deformations. Hence, all dependence from moduli orthogonal to the volume comes from the Casimir energy. In general, the $(d+q)$-dimensional potential will take the schematic form
\begin{equation}\label{eq:11d.potential}
    V_{d+q}(R,h) = \frac{\ell_{\text{Pl},d+q}^{q-d}}{2R^{2q}}\left ( \frac{N}{2\pi} \right )^2 - \frac{2 v(h)}{R^{d+q}} \, ,
\end{equation}
where $v(h)$ is a function of the moduli orthogonal to the volume $h$ that can be computed using the methods described in \cref{sec:casimir_torus_pert}. The original vacuum solution is chosen to be at the origin of the (pseudo-)moduli space, so that $v(0)=|\rho_c|$. From the considerations in \cref{sec:tadpoles} it also follows that $v'(0) = 0$, since, the solution is on-shell along all directions.

The masses of the fields $h$, $R$, can be inferred from the lower-dimensional potential in the Einstein frame. Denoting by $R = R_{*}$ the stabilized value of the internal radius, the following metric ansatz provides the change of frame,
\begin{equation}\label{eq:metric.einstein}
    ds^2 = \left (\frac{R_{*}}{R} \right )^{\frac{2q}{d-2}} L ^2 ds^2_{\text{AdS}_d} + R^2 ds_{T^q} \, .
\end{equation}
After integrating over the torus, the dimensionally reduced energy density reads
\begin{equation}\label{eq:4d.potential}
    V_{d} = (2\pi R_*)^q R_*^{\frac{2q}{d-2}}  \left (\frac{\ell_{\text{Pl},d+q}^{q-d}}{2R^{2q\frac{d-1}{d-2}}}\left ( \frac{N}{2\pi} \right )^2 - \frac{2v(h)}{R^{d\frac{d+q-2}{d-2}}} \right) .
\end{equation}
Provided that $q>d>1$, this potential always has a minimum in $R$, since the first term always scales with a higher power of the inverse radius than the second. In particular, for $h=0$ it is located at
\begin{equation}
    R_* = \ell_{\text{Pl},d+q} \left ( \frac{N}{2\pi}\right )^{\frac{2}{q-d}} v(0)^{-\frac{1}{q-d}} \left ( \frac{q(d-1)}{2d(d+q-2)}  \right )^{\frac{1}{q-d}} \, .
\end{equation}
This agrees with the higher-dimensional computation in \cref{eq:radii}. From \cref{eq:4d.potential} it is not obvious that $h = 0$ will be a minimum, or more generally that its perturbations will satisfy the \ac{BF} bound. We shall now address this question in detail.

Before doing so, note that in this discussion it is essential that the fields be canonically normalized, so that their masses can be directly extracted from the potential. To achieve this, we perform the field redefinitions
\begin{equation}\label{eq:redefinitions}
    h = a\ell_{\text{Pl},d}^{\frac{d-2}{2}} \phi \, , \quad R = R_{*}e^{b\ell_{\text{Pl},d}^{\frac{d-2}{2}}\rho} \, ,
\end{equation}
where $a$ and $b$ are numerical factors such that the kinetic term of the fields in the dimensionally reduced action appear with the canonical normalization (e.g. $1/2 (\partial \rho)^2$). For the redefinition of $h$, we have assumed that the kinetic term of $h$ is already proportional to $(\partial h)^2$ in the action. For simplicity, in the following we use a notation for which $\phi$ comprises a single modulus. In the examples below we focus on this case, since finding a single tachyonic mode below the \ac{BF} bound is enough to conclude the presence of perturbative instabilities.

The mass spectrum of this vacuum can then be determined by evaluating the Hessian of the potential at the point $(\rho,\phi)=(0,0)$. Since $v'(0)=0$, off-diagonal entries of the mass matrix vanish, and the physical masses of the fields can be directly read off from the second derivatives of the potential. As we have noted around \cref{eq:4d.potential}, the mass of $\rho$ is positive. However, other (pseudo-)moduli could have negative squared masses, since
\begin{equation}
    m_{\phi}^2=\frac{\partial^2 V_d}{\partial\phi^2} =-\ell^{d+q-2}_{\text{Pl},d+q} \frac{2a^2 v''(0)}{R_*^{d+q}}
\end{equation}
which is negative if $v''(0)>0$. Nevertheless, since we are working in \ac{AdS}, the true tachyonic directions are those violating the \ac{BF} bound \cite{Breitenlohner:1982jf}. On an \ac{AdS} spacetime of radius $L$, the \ac{BF} bound for a scalar field of mass $m$ reads
\begin{equation}\label{eq:BF.bound}
    m^2L^2 \geq -\frac{(d-1)^2}{4} \, .
\end{equation}
Recalling the explicit form of $L$ in \cref{eq:radii}, we find
\begin{equation}
    m_{\phi}^2L^2 = -2 a^2 \frac{q(d-1)^2}{q-d} \frac{v''(0)}{v(0)} \, ,
\end{equation}
We see that the quantity $m_{\phi}^2 L^2$ is generically of order one, and must be evaluated on a case-by-case basis to determine whether true tachyonic modes appear in the spectrum.

Let us particularize to our case of interest $d=4$, $q=7$. We consider the following deformation of the torus orthogonal to the volume
\begin{equation}\label{eq:deformation}
    R^2 ds_{T^7}^2= R_*^2 e^{2b\rho} \left (e^{2a \phi} dy_1^2+e^{-2a \phi} dy_2^2 + \sum_{i=3}^{7} dy_i^2 \right ),
\end{equation}
where $a$ and $b$ are the constants from \cref{eq:redefinitions} in Planck units. The first step is to find their precise values. Using the metric ansatz in \cref{eq:metric.einstein} with the metric on the torus given by \cref{eq:deformation}, the eleven-dimensional Einstein-Hilbert term decomposes according to as\footnote{In the following expressions we have rescaled the Planck lengths to include the factors of $2\pi$ used in \cref{sec:casimir_vacua} to match the conventions of \cite{Luca:2022inb}. This does not affect the ensuing discussion.}
\begin{equation}\label{eq:eh.action}
   \frac{1}{\ell_{\text{Pl},11}^9} \int d^{11} x \sqrt{-g_{11}} \mathcal{R}_{11} = \frac{1}{\ell_{\text{Pl},4}^2} \int d^{4} x  \sqrt{-g_4} \left (\mathcal{R}_4 - \frac{63}{2}b^2(\nabla \rho )^2-2a^2(\nabla \phi )^2 \right ).
\end{equation}
Imposing that the kinetic terms be canonical, this sets
\begin{equation}\label{eq:coefficients}
    a = \frac{\ell_{\text{Pl},4}}{2} \, , \quad b = \sqrt{\frac{1}{63}} \,  \ell_{\text{Pl},4} \, .
\end{equation}
The Casimir term in the potential takes the form
\begin{equation}\label{eq:potential.casimir.bit}
    \ell_{\text{Pl},4}^{2}V_{\text{Cas},4d}(\rho=0,\phi) = - 42\,\frac{v(2a\phi)}{v(0)}\frac{1}{L^2} \, ,
\end{equation}
with the function $v(x)$ given by
\begin{equation}\label{eq:deformation_v}
    \frac{v(x)}{v(0)} = 
    \dfrac{
        \displaystyle\int_0^\infty \! \frac{ds}{s^{3}} \left[
            \theta_3\!\left(e^{-e^x s}\right)
            \theta_3\!\left(e^{-e^{-x}s}\right)
            \theta_3\!\left(e^{-s}\right)^{5}
            -
            \theta_2\!\left(e^{-e^x s}\right)
            \theta_2\!\left(e^{-e^{-x}s}\right)
            \theta_2\!\left(e^{-s}\right)^{5}
        \right]
    }{
        \displaystyle\int_0^\infty \! \frac{ds}{s^{3}} \left[
            \theta_3\!\left(e^{-s}\right)^{7}
            -
            \theta_2\!\left(e^{-s}\right)^{7}
        \right]
    } \, .
\end{equation}
To evaluate the second derivative of this expression, we numerically computed the integral for values of $\phi$ near zero and then approximated the second derivative using a finite-difference formula. Concretely, we used
\begin{equation}\label{eq:finite.differences}
    v''_\epsilon(0) \approx \frac{v(\epsilon)-2v(0)+v(-\epsilon)}{\epsilon^2} \, ,
\end{equation}
where $\epsilon$ is a small real number. Taking $\epsilon=10^{-5}$, we find that the second derivative is positive,
\begin{equation}
    \frac{v''(0)}{v(0)}\approx8.160 \, .
\end{equation}
As a result, a sufficiently precise computation of its value is required in order to determine whether $\phi$ is tachyonic. Using \cref{eq:potential.casimir.bit}, the mass is found to be
\begin{equation}\label{eq:first_BF_bound}
    m_{\phi}^2L^2 = -168 \,\frac{a^2}{\ell_{\text{Pl},d}^2} \frac{v''(0)}{v(0)} = -42 \,\frac{v''(0)}{v(0)} \approx -342.73 \, ,
\end{equation}
which is well below the bound in \cref{eq:BF.bound}. This shows that this family of parametrically scale-separated solutions suffers from perturbative instabilities.

A toy model in which the possibility of tachyonic modes can be seen more explicitly is the two-torus, for which the function $v$ of the complex structure $\tau = \tau_1 + i \tau_2$ is modular-invariant \cite{Cheng:2001ti, Ito:2003tc, Oikonomou:2009pg, Luo:2022tqy}. Specifically, $v \propto E_{-\frac{d}{2}}(\tau)$ is proportional to a real-analytic Eisenstein series \cite{Luo:2022tqy}, where the square torus corresponds to $\tau = i$. The Laplacian equation 
\begin{eqaed}\label{eq:eisenstein_eq}
    -\tau_2^2(\partial_{\tau_1}^2 + \partial_{\tau_2}^2) E_s = s(1-s) E_s
\end{eqaed}
then implies that the trace of the Hessian matrix, normalized by $v(0) \propto E_{-\frac{d}{2}}(i)$, is given by
\begin{eqaed}\label{eq:hess_trace_T2}
    \frac{(\text{tr Hess }v)(0)}{v(0)} = \frac{2\tau_2^2 \, (\partial_i \partial^i E_{-\frac{d}{2}})(i)}{E_{-\frac{d}{2}}(i)} = d\left(\frac{d}{2}+1\right) \, ,
\end{eqaed}
where the factor of $\tau_2^2$ arises contracting the partial derivatives with the metric tensor of the complex-structure field space. Hence, there is a positive eigenvalue at least as large as $\frac{d}{2}\left(\frac{d}{2} + 1\right) = 1$ for $d = 4$. If we could plug this into \cref{eq:first_BF_bound} directly we would conclude that tachyons are present. However, since $q>d$ for the solution to exist, we need to embed the two-torus inside the full $q$-torus, as in \cref{eq:deformation_v}.

As an aside, there is another solution with $d>3$ extended spacetime dimensions that could be embedded in eleven-dimensional gravity with a form field. This corresponds to a family of scale separated $\text{AdS}_5\times T^6$ vacua. In that case, the Casimir energy is given by
\begin{equation}
     \ell_{\text{Pl},5}^{3}V_{\text{Cas},5d}(\rho=0,\phi) = - 192\,\frac{v(2a\phi)}{v(0)}\frac{1}{L^2} \, ,
\end{equation}
and the second derivative of $v(x)$ is now
\begin{equation}
    \frac{v''(0)}{v(0)} \approx  9.735 \, .
\end{equation}
The constant $a$ stays the same (in Planck units) and, when the dust settles, one obtains
\begin{equation}
    m_{\phi}^2 L^2 \approx -1.87 \times 10^{3} \, ,
\end{equation}
which is again below the \ac{BF} bound. Once more, we remark that this bottom-up solution does not account for potential gravitating \ac{UV}-sensitive terms in the vacuum energy, which are generically present since this eleven-dimensional gravity with a five-form flux is not a truncation of any supergravity.

We conclude the analysis with a comment on kinetic terms. The Casimir energy modifies the potential term of the effective action, but quantum effects in general also modify derivative terms via form factors. One could ask whether the resulting quantum corrections to the masses of fluctuations could change the above considerations. Schematically, a form factor arising from quantum corrections would correct the $k^2$ structure of kinetic terms by terms of the form $\frac{1}{L^2} f(R^2k^2)$. Parametrizing $f(x) \sim x^n$ with a power law, we have $n>1$ for genuine corrections to $k^2$. This corrections becomes significant for $L \abs{k} \sim R^n \abs{k}^n$, which leads to
\begin{eqaed}\label{eq:threshold_k}
    R\abs{k} \sim N^{\frac{3}{n-1}} \gg 1 \, .
\end{eqaed}
Therefore, no significant deviations to the above analysis occurs within the (would-be) four-dimensional \ac{EFT} in which we found violations of the \ac{BF} bound.

\subsection{Non-perturbative instabilities}\label{sec:general_decay}

The scale-separated solutions at stake are entirely specified by a single flux quantum $N$. In the limit of large flux, scale separation is achieved and the curvature scales become parametrically small, allowing for a controlled semiclassical analysis. Even in the absence of classical instabilities, non-supersymmetric AdS vacua are expected to decay via flux tunneling \cite{Horowitz:2007pr, Brown:2010bc, Brown:2010mf, Brown:2010mg, Ooguri:2016pdq, Antonelli:2019nar, Dibitetto:2020csn, Bomans:2021ara, Dibitetto:2022rzy}, unless a positive-energy theorem holds \cite{Giri:2021eob, Menet:2025nbf} (possibly along the lines of fake supersymmetry \cite{Raucci:2023xgx, Raucci:2024fnp}). This process reduces the flux number, eventually reaching a regime in which curvature scales become large and the semiclassical approximation breaks down.

Let us briefly review how to compute the decay rate for flux tunneling in the semiclassical regime. The decay rate per unit volume is proportional to $e^{-S^{E}}$, where $S^E$ is the euclidean action of the brane in consideration. The exponential is the dominant term in the sense that the decay rate is logarithmically asymptotic to it. We consider a $(d-2)$-brane instanton mediating a decay of \ac{AdS}$_d$. Its Euclidean action is given by
\begin{equation}
    S^E = T \int d^{d-1} x \sqrt{j^{*}g} - \mu \int C_{d-1} = S^{E}_{\text{Area}} - S^{E}_{\text{Vol}},
\end{equation}
where $j$ is the embedding of the brane in spacetime, $T$ is the tension of the brane and $\mu$ its charge. Since we ultimately consider M2-branes in M-theory, no dilaton field needs to be included. The first term in the action is the area swept by the worldvolume, which for a spherical brane of radius $\rho$ reads
\begin{equation}
    S^{E}_{\text{Area}} = T \Omega_{d-1}\rho^{d-1} \, .
\end{equation}
The second term can be computed using Stokes' theorem. Denoting the volume enclosed by the brane as $V$, one has
\begin{equation}
    S^{E}_{\text{Vol}} = \mu \int_{\partial{V}} C_{d-1} = \mu \int_{V} dC_{d-1} = \mu \int_V F_{d} \, .
\end{equation}
Recall that, in our case, the electric flux $F_d$ threads the whole \ac{AdS}$_d$ space. Concretely, it takes the form
\begin{equation}
    F_{d} = f_d \, \text{vol}_{\text{AdS}_d}\,, \quad f_d = \frac{N}{2\pi} \ell_{\text{Pl},{d+p}}^{q-1} \frac{L^d}{R^q} \,,
\end{equation}
after applying the quantization condition analogous to \cref{eq:quantization_condition}. Here, vol$_{\text{AdS}}$ is the volume form of an $\text{AdS}_d$ space with unit radius. In order to compute the volume integral, the (Euclidean) \ac{AdS} metric can be parametrized as
\begin{equation}
    ds^2_E = d\xi^2 + \tilde\rho^2(\xi)d\Omega_{d-1}^2 \, ,
\end{equation}
where $\tilde\rho$ obeys
\begin{equation}
    (\tilde\rho')^2 = 1 + \tilde\rho^2 \, .
\end{equation}
For a spherical brane of radius $\rho$, the integral over the volume yields
\begin{equation}
    S_{\text{Vol}}^E = \mu f_d \int_V \text{vol}_{\text{AdS}_d} = \mu \Omega_{d-1}f_d \int d \xi \tilde\rho^{d-1}(\xi) = \mu \Omega_{d-1}f_d\int_0^{\frac{\rho}{L}} d\tilde \rho \frac{\tilde \rho^{d-1}}{\sqrt{1+\tilde\rho^2}} \, .
\end{equation}
Defining
\begin{eqaed}
    \mathcal{V}(x) &= \int_0^{\frac{\rho}{L}} d\tilde \rho \frac{\tilde \rho^{d-1}}{\sqrt{1+\tilde\rho^2}} \, , \\ x &= \frac{\rho}{L} \, ,  \\ \beta &= \frac{f_d}{(d-1)L^{d-1}}\frac{\mu}{T} \, ,
\end{eqaed}
the Euclidean action takes the form \cite{Antonelli:2019nar}
\begin{equation}\label{eq:euclidean.action}
    S^{E} = T \Omega_{d-1} L^{d-1} \left ( x^{d-1} - (d-1)\beta \,\mathcal{V}(x)\right ).
\end{equation}
Analyzing this expression, it follows that if $\beta > 1$, the action admits a maximum; whereas if $\beta < 1$, the action is unbounded from above. Since this expression appears in the exponent of the decay rate formula, brane nucleation will be suppressed in the latter case. Moreover, $\beta$ must be $\mathcal{O}(N^0)$ for the semiclassical computation to be consistent.

We have verified that indeed $\beta > 1$ and that it is of order one in the large-flux limit, indicating that this is an allowed non-perturbative decay channel. In particular, this applies to the original $d=4$, $q=7$ vacuum, where $\beta = 2\sqrt{2}$. In general,
\begin{equation}
    \beta(d,q) = \sqrt{\frac{2d(d+q-2)}{(d-1)(q-d)}} > 1 \, .
\end{equation}
It is also interesting to compute the parametric scaling of the decay rate. Concretely, we find that $S^E \sim N^\frac{(d+q)(d-1)}{q-d}$, which for our case of interest reads $S^E \sim N^{11}$. This is a higher suppression than what one finds in other non-supersymmetric heterotic and orientifold models \cite{Antonelli:2019nar}, where no parametric scale-separation is present. 

\section{Conclusions}\label{sec:conclusions}

Flux compactifications supported by Casimir energy provide a simple setting to seek string vacua with parametric scale separation. Moreover, they naturally incorporate supersymmetry breaking, another essential requirement for realistic models. This feature prompted a search for de Sitter solutions of this type \cite{ValeixoBento:2025yhz}, although they cannot be neither minima nor parametrically scale-separated.

In this paper we have shown that the simplest construction of this type, where the internal space is a square torus, is unstable. While perturbative instabilities, in the form of Breitenlohner-Freedman tachyons, may be avoidable in more elaborate settings, non-perturbative instabilities (here in the form of M2-brane nucleation) are more robust, since they do not depend on the internal geometry. These lessons mirror the findings of \cite{Basile:2018irz, Antonelli:2019nar, Baykara:2022cwj} where flux compactifications on curved internal manifolds are supported by non-supersymmetric dilaton potentials. The presence of nucleation instabilities prevents the existence of a CFT dual \cite{Ooguri:2016pdq}, since the vacuum decays immediately close to the conformal boundary. Therefore, a holographic interpretation presumably involves a renormalization group flow \cite{Antonelli:2018qwz, Ghosh:2021lua, Banerjee:2023uto}. Nevertheless, low-energy observers in the bulk can survive up to an AdS time, which is parametrically larger than the cutoff of the \ac{EFT}. In contrast, in the presence of perturbative instabilities the dimensionally reduced \ac{EFT} lacks even a perturbative vacuum, and is ill-defined. In this sense, even if brane nucleation is unavoidable, for the purposes of scale separation it would be sufficient to find a solution devoid of Breitenlohner-Freedman tachyons. Combining the methods developed in this paper with those of \cite{ValeixoBento:2025yhz, DallAgata:2025jii} would allow for a vast selection of compactification manifolds, whose reduced number of classical moduli may leave room for perturbatively stable vacua. We leave this task for future work. Another possibility is to combine the Casimir energy with stringy ingredients such as orientifolds, in the spirit of \cite{DeWolfe:2005uu, Caviezel:2008ik, Farakos:2020phe, Cribiori:2021djm, Apers:2022zjx, Farakos:2023nms, Tringas:2023vzn, Andriot:2023fss, VanHemelryck:2024bas, VanHemelryck:2025qok, Tringas:2023vzn, Carrasco:2023hta, Arboleya:2024vnp, Arboleya:2025ocb}, whose subtleties involving uplifts or backreaction are an active topic of research \cite{Banks:2006hg, Gautason:2015tig, Junghans:2020acz, Marchesano:2020qvg, Junghans:2023yue, Coudarchet:2023mfs}.

It may be the case that no parametric scale separation is possible in the string landscape, whether with a geometric internal manifold or otherwise non-geometric mass gap to new physics. In the context of stable asymptotically AdS sectors, this may be a consequence of holography \cite{Collins:2022nux, Lust:2022lfc, Bobev:2023dwx, Bena:2024are} or other swampland constraints \cite{Cribiori:2022trc, Cribiori:2022sxf, Montero:2022ghl, Cribiori:2023gcy, Cribiori:2023ihv, Cribiori:2024jwq, Montero:2024qtz}. More generally, if the set of nonequivalent \acp{EFT} in the landscape is finite \cite{Hamada:2021yxy, Delgado:2024skw, Kim:2024eoa} regardless of the cutoff, scale separation may only be achievable in a numerical sense. This would be embodied in the constructions of \cite{Demirtas:2021nlu, Demirtas:2021ote, McAllister:2024lnt, Abel:2024vov, ValeixoBento:2025yhz}; strictly speaking, this is the only strict requirement for phenomenology, insofar as one is willing to relinquish formal control to an extent. Alternatively, outside the context of (metastable)stable vacua, the parameter realizing scale separation could be replaced by time evolution \cite{Andriot:2025cyi}. This may be a phenomenologically favored approach \cite{Alestas:2024gxe, Casas:2024xqy, Andriot:2024jsh, Andriot:2024sif, ValeixoBento:2025yhz, Bedroya:2025fwh} in light of recent experimental evidence \cite{DES:2024jxu, DESI:2025fii, DESI:2025zgx}. A different construction that also hinges on time dependence is embodied by the dark bubble scenario \cite{Banerjee:2018qey, Banerjee_2019, Banerjee:2020wix,Banerjee:2020wov,Banerjee:2021yrb,Danielsson:2021tyb,Banerjee:2022ree,Danielsson:2022lsl,Danielsson:2022fhd, Danielsson:2023alz, Basile:2023tvh, Banerjee:2023uto,danielsson2024chargednariaiblackholes}, which replaces four-dimensional \acp{EFT} with an expanding brane-world. Scale separation is achieved in a different fashion, but the realizations proposed thus far predict significant deviations in low-energy physics \cite{Basile:2025lwx} and cosmology \cite{Basile:2025lek}.

Lastly, Minkowski vacua with no moduli would technically realize infinite scale separation without any small control parameter \cite{Rajaguru:2024emw, Chen:2025rkb}. These settings may provide viable avenues to realize more realistic models within genuinely four-dimensional \acp{EFT}. In perturbative string theory this can almost be achieved, leaving only the dilaton as neutral (pseudo-)modulus, via non-geometric constructions of \emph{CFT islands} \cite{Baykara:2023plc, DeFreitas:2024ztt, Angelantonj:2024jtu, Leone:2024hnr, Baykara:2024tjr, Baykara:2025lhl, Aldazabal:2025zht}. Among these, the non-supersymmetric ones generate a runaway dilaton potential. The methods of \cite{Mourad:2016xbk} can be adapted to seek Freund-Rubin vacua with parametric control over higher-derivative corrections and string loops, although instabilities would still lurk around the corner \cite{Basile:2018irz, Antonelli:2019nar}.

\section*{Acknowledgements}\label{sec:acknowledgements}

The authors thank Bruno V. Bento, Elias Kiritsis, Bruno De Luca, Luca Martucci, Miguel Montero, Salvatore Raucci, Alessandro Tomasiello and Thomas Van Riet for useful discussions. N.R. was supported in part by MIUR-PRIN contract 2022YZ5BA2  ``Effective quantum gravity"  and in part by MIUR-PRIN contract 2017CC72MK003. The work of I.B. is supported by the Origins Excellence Cluster and the German-Israel-Project (DIP) on Holography and the Swampland.

\printbibliography

\end{document}